\documentclass[fleqn,usenatbib]{mnras}

\usepackage{newtxtext,newtxmath}

\usepackage[T1]{fontenc}
\usepackage{ae,aecompl}

\usepackage{graphicx}	
\usepackage{amsmath}	
\usepackage{bm}
\usepackage[usenames]{color}
\usepackage{animate}
\usepackage{mathrsfs}

\graphicspath{{Figures/}}
\usepackage[normalem]{ulem}
\usepackage{array}
\usepackage{framed}

\title[regimes in lensing]{Regimes in astrophysical lensing: refractive optics, diffractive optics, and the Fresnel scale}

\author[Jow et al.]{
Dylan L. Jow,$^{2,3}$\thanks{E-mail: djow@physics.utoronto.ca}
Ue-Li Pen,$^{1,2,3,4,5,6}$
and Job Feldbrugge$^{7,8}$
\\
$^{1}$Institute of Astronomy and Astrophysics, Academia Sinica, Astronomy-Mathematics Building, No. 1, Section 4,
Roosevelt Road, Taipei 10617, Taiwan \\
$^{2}$Canadian Institute for Theoretical Astrophysics, University of Toronto, 60 St. George Street, Toronto, ON M5S 3H8, Canada\\
$^{3}$Department of Physics, University of Toronto, 60 St. George Street, Toronto, ON M5S 1A7, Canada\\
$^{4}$Perimeter Institute for Theoretical Physics, 31 Caroline St. North, Waterloo, ON, Canada N2L 2Y5\\
$^{5}$Canadian Institute for Advanced Research, CIFAR program in Gravitation and Cosmology\\
$^{6}$Dunlap Institute for Astronomy \& Astrophysics, University of Toronto, AB 120-50 St. George Street, Toronto, ON M5S 3H4, Canada\\
$^{7}$Higgs Centre for Theoretical Physics, James Clerk Maxwell Building, Edinburgh EH9 3FD, UK\\
$^{8}$Department of Physics, Carnegie Mellon University, 5000 Forbes Ave, Pittsburgh, PA 15217, USA
}

\date{Accepted XXX. Received YYY; in original form ZZZ}

\pubyear{2021}

\begin{document}
\label{firstpage}
\pagerange{\pageref{firstpage}--\pageref{lastpage}}
\maketitle

\begin{abstract}
 Astrophysical lensing has typically been studied in two regimes: diffractive optics and refractive optics. Previously, it has been assumed that the Fresnel scale, $R_F$, is the relevant physical scale that separates these two regimes. With the recent introduction of Picard-Lefschetz theory to the field of lensing, it has become possible to generalize the refractive description of discrete images to all wave parameters, and, in particular, exactly evaluate the diffraction integral at all frequencies. In this work, we assess the regimes of validity of refractive and diffractive approximations for a simple one-dimensional lens model through comparison with this exact evaluation. We find that, contrary to previous assumptions, the true separation scale between these regimes is given by $R_F/\sqrt{\kappa}$, where $\kappa$ is the convergence of the lens. Thus, when the lens is strong, refractive optics can hold for arbitrarily small scales. We also argue that intensity variations in diffractive optics are generically small, which has implications for the study of strong diffractive scintillation (DISS).
\end{abstract}

\begin{keywords}
waves -- radio continuum: ISM -- pulsars:general -- fast radio bursts 
\end{keywords}



\section{Introduction}
\label{sec:intro}

When coherent sources, such as pulsars and fast radio bursts (FRBs), undergo lensing, wave effects become important. Pulsar scintillation, for example, is due to the interference of many coherent images formed due to scattering in the interstellar medium (ISM). As more data of these coherent sources becomes available from current and future observations, there has been increased interest in the study study of wave optics in the context of astrophysical lensing \citep[see e.g.][]{GrilloCordes2018, 2018Natur.557..522M, job_pl, 2020arXiv200801154F, 2020arXiv201003089F, Jow2020, 2021MNRAS.506.6039S}. 

In principle, wave optics is fully described by the Kirchhoff-Fresnel diffraction integral, which is a path integral over all paths connecting the source and observer through a refractive medium. In practice, however, the full diffraction integral is rarely computed, as there exist few analytic solutions for lenses of interest, and its highly oscillatory nature makes it challenging to compute numerically.  It has only been through the recent application of the mathematical framework of Picard-Lefschetz theory that computing the diffraction integral exactly has become numerically tractable \citep[see e.g.][]{job_pl,2020arXiv200801154F, 2020arXiv201003089F, Jow2021,2021MNRAS.506.6039S}. Previously, astrophysical lensing has typically been studied in the geometric regime, where the contribution to the observed flux is isolated to a discrete set of well-defined images whose individual fluxes can be easily computed. Wave effects are then added, to first order, by computing the phase associated with each image and allowing them to coherently interfere at the observer. This is also known as the Eikonal limit, which is equivalent to the stationary-phase approximation of the diffraction integral\footnote{Note that some authors make a distinction between geometric optics and the Eikonal limit of wave optics, where geometric optics is when the images are added incoherently at the observer and the Eikonal limit is when they are added coherently. Here we will use the terms ``geometric optics" and ``Eikonal limit" interchangeably to refer to the regime where isolated images interfere coherently.}. 

Geometric optics, or the Eikonal limit, is the high-frequency limit of the Kirchhoff-Fresnel integral. The low-frequency limit of optics has also been studied, as it is often possible to compute the Kirchhoff-Fresnel integral through a perturbative expansion. It is of central importance to the study of wave optics to understand at what frequencies these two regimes become valid, and whether or not there is any overlap. In the absence of an exact result for the Kirchhoff-Fresnel integral, it has often been assumed in the literature that the Fresnel scale sets the frequency which separates geometric optics from the full wave regime \cite[see e.g.][]{1986ApJ...310..737C, Fiedler1987, 1993Natur.366..320C, 1998ApJ...496..253C, 2012MNRAS.421L.132P,2018ApJ...865..104J, 2018MNRAS.474.4637K,dong_extreme_2018, 2022MNRAS.509.5872E}. In this work, however, we will show that the Fresnel scale is, generically, the wrong scale. Rather, a combination of the Fresnel scale and lens strength sets the correct separation scale between the geometric and full wave regimes.

The results presented in this paper are relevant to the study of the lensing of coherent sources broadly speaking, including both gravitational and plasma lensing. However, in particular, our results bear implications for the study of the scintillation of coherent radio sources (e.g. pulsars) due to scattering in the ISM. The geometric and wave regimes of interstellar scintillation have been studied in detail using approximations and heuristic arguments since the discovery of pulsar scintillation \citep{1975ApJ...196..695L, 1977ARA&A..15..479R}. In this literature, the geometric and full-wave regimes go by the names refractive interstellar scintillation (RISS) and diffractive interstellar scintillation (DISS), respectively. We will argue that the usage of the terms ``refractive" and ``diffractive" to describe these regimes is misleading, and that, in fact, most if not all scintillation can be described by the stationary-phase approximation, which we identify with refractive optics. This is not merely a semantic preference, but has important implications for the physical inferences one may draw from observations, as we will discuss in Section~\ref{sec:turbulence}.

It is by employing the mathematical framework of Picard-Lefschetz theory that we are able to make a detailed study of the transition from geometric optics to full, wave optics. Not only does Picard-Lefschetz theory allow for a non-perturbative, exact evaluation of the Kirchhoff-Fresnel integral at all frequencies, it is also a conceptually powerful framework for analyzing problems in optics. As we discuss in Appendix~\ref{sec:appendixB}, Picard-Lefshcetz theory allows one to make a well-defined separation of the total Kirchhoff-Fresnel integral into contributions from a discrete set of images, which are simply the classical images of geometric optics. This separation is well-defined at arbitrary frequencies, even deep into the diffractive / perturbative regime where previously it has been assumed that the notion of an optical ``image" breaks down. Indeed, as shown in Appendix~\ref{sec:appendixB}, Picard-Lefschetz theory allows us to track the contributions of each image as one goes from the geometric regime to the perturbative regime, thus allowing us to bridge the conceptual gap between refractive and diffractive optics. Thus, in addition to solving a decades old numerical problem, Picard-Lefschetz theory represents a genuine step forward in the theory of optics.  It is within the context of these newfound insights into the mathematical structure of optics (discussed further in \citet{job_pl, Jow2021} that we put forward our arguments in this paper.

This paper is structured in the following way. In Section~\ref{sec:pert_eik}, we will review the stationary-phase approximation which is used to compute geometric optics, as well as the perturbative expansion which can be used to approximate the diffraction integral in the low-frequency limit. In Section~\ref{sec:ratlens}, we will study a simple 1D lens with a simple rational potential, $\psi(x) \propto 1 / (1 + x^2)$, to argue that most of parameter space is well described by either of these two approximations, and that the separation scale between the two regimes is given by the Fresnel scale divided by the square-root of the convergence. In Section~\ref{sec:definitions} we will discuss the usage of the words ``diffraction", ``refraction", and ``interference", as these words often attain different connotations and meanings in different sub-fields, and indeed, between different authors within those fields. Based on these definitions given, we will identify these two approximations with diffractive and refractive optics. Thus, contrary to what has previously been assumed in much of the lensing literature, the separation scale between refractive and diffractive optics is \textit{not} the Fresnel scale alone. Indeed, for either large or small lens strengths, the Fresnel scale may be off by orders-of-magnitude. In Section~\ref{sec:strong}, we will argue that lensing in the diffractive regime is typically weak, and therefore an observation of strong scintillation already suggests lensing in the refractive regime. In Section~\ref{sec:tdelays} we show how the two regimes behave differently in delay space. Our numerical results are quite general and can be applied to scenarios in gravitational lensing and plasma lensing, as well as potential applications to scattering theory in quantum mechanics. However, in Section~\ref{sec:turbulence}, we discuss in greater detail the connection between our results and previous results in the literature on turbulent plasma lensing. In Appendix~\ref{sec:appendixB}, we present the detailed Picard-Lefschetz analysis of the rational lens.

We note that while our focus will be astrophysical lensing, our results may be applicable to problems in scattering theory in quantum mechanics. In particular, while we focus on single-plane lensing, the formalism for multi-plane lensing is equivalent to the real-time Feynman path integral in the limit of infinite planes using the time-slicing approximation \citep{2020arXiv201003089F}. In this setting, the diffractive limit corresponds to the application of perturbative methods resulting in the Feynman rules, whereas the refractive limit corresponds to non-perturbative methods such as the use of instantons and the study of (complex) classical trajectories. The tools of Picard-Lefschetz theory, which we use here to evaluate the Kirchoff-Fresnel integral, have been recently used to evaluate the Feynman path integral to re-interpret phenomena in quantum mechanics \citep{2014NJPh...16f3006T, 2014arXiv1408.0012C, 2014AnPhy.351..250T, 2021AnPhy.42968457G}.

\section{The perturbative expansion and Eikonal approximation of wave optics}
\label{sec:pert_eik}

\begin{figure}
    \centering
    \includegraphics[width=\columnwidth]{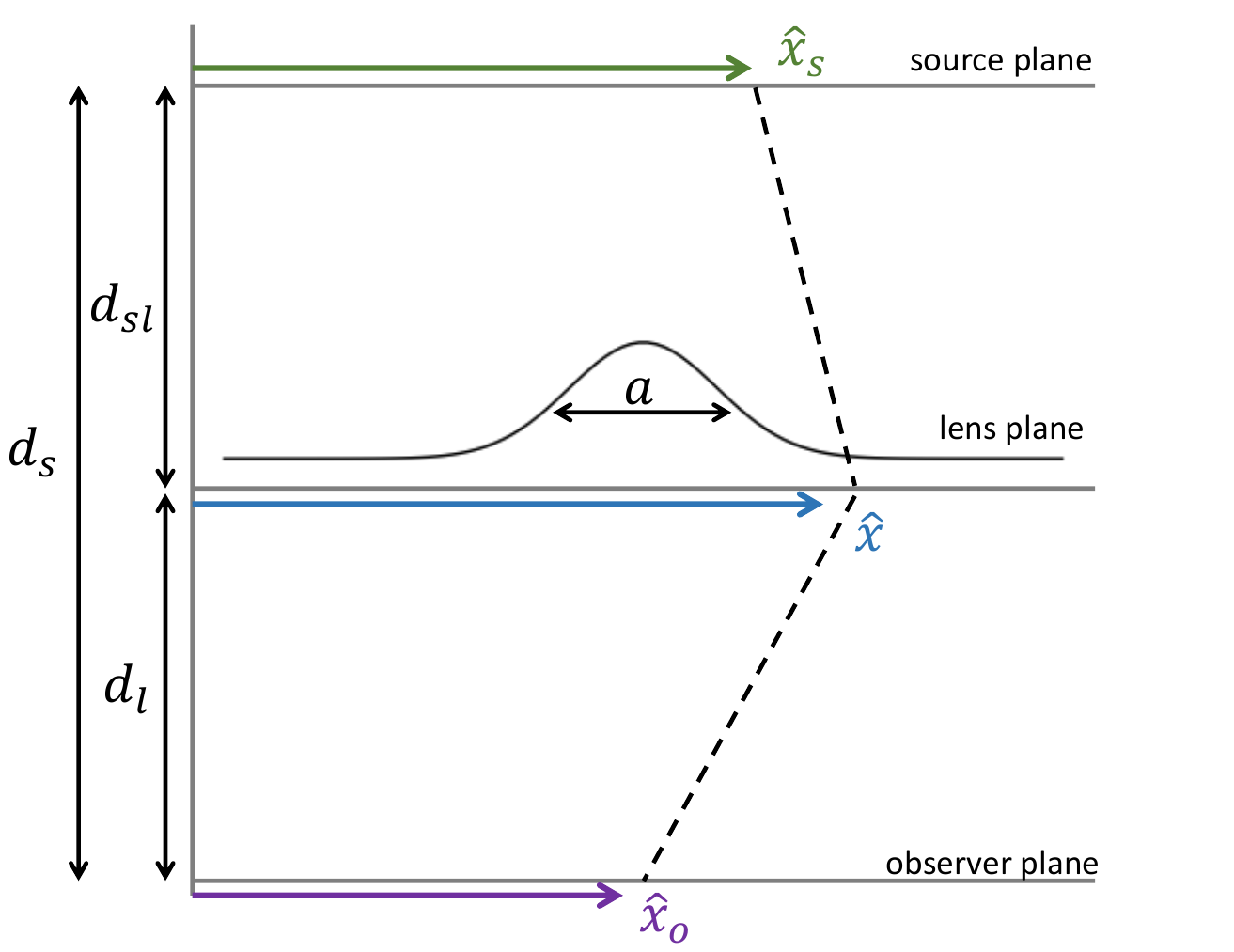}
    \caption{Diagram of thin-plasma lensing showing distances and co-ordinates involved.}
    \label{fig:lensing_geometry}
\end{figure}

The flux at $y$ of a source lensed by a thin screen can be expressed as the magnitude $|F(y)|^2$, where the amplitude $F$ is obtained from the dimensionless Kirchhoff-Fresnel diffraction integral
\begin{equation}
F(y) = \sqrt{\frac{\nu}{2\pi i}} \int_{-\infty}^\infty e^{i\nu [ \frac{(x-y)^2}{2} + \kappa \psi(x) ]} dx\,,
    \label{eq:Kirchhoff-Fresnel}
\end{equation}
with the phase variation induced by the lens given by $\psi$ and where we integrate over all possible paths from the source to the observer intersecting the lens plane at $x$. For convenience, we use the dimensionless version of the Kirchhoff-Fresnel integral with the dimensionless relative position of source and observer, frequency, and lens strength $y$, $\nu$, and $\kappa$. Fig.~\ref{fig:lensing_geometry} shows a diagram of the lensing setup, with dimensionful parameters. The dimensionless parameters in Eq.~\eqref{eq:Kirchhoff-Fresnel} are related to these by $x = \hat{x}/a$, where $a$ is some length scale associated with the lens, $y = \hat{y}/a = (\hat{x}_s d_l + \hat{x}_o d_{sl}) / a d_s $, and $\nu = \omega a^2 / c \overline{d}$ where $\overline{d} = d_{sl} d_l / d_s$. In the following sections, we will refer only to the dimensionless parameters so that our analysis can be easily applied to a wide variety of contexts. We will further discuss the underlying physical parameters in the context of plasma and gravitational lensing in Sections \ref{sec:turbulence} and \ref{sec:values}. See also \citet{Schneider}, \citet{nakamura}, \citet{job_pl} and \cite{Jow2021} for a more thorough discussion of the correspondence of these dimensionless parameters to physical parameters in different contexts.

In this work, we restrict our attention to the case of a bounded, one-dimensional lens, with a single peak, and we make a choice of co-ordinates $x$ and $y$ such that the peak is centred at $x=0$ and has curvature $\psi''(0) = -2$. Under this choice of coordinates, the parameter $\kappa$ is the convergence of the lens \citep{2012MNRAS.421L.132P, Schneider}. A canonical example of such a lens is the Gaussian lens, $\psi(x) = \exp\{-x^2\}$, which has been used in many lensing studies \citep[see, e.g.,][]{1998ApJ...496..253C,2012MNRAS.421L.132P,2017ApJ...842...35C}. One may also construct rational approximations of the Gaussian lens, such as the simple rational lens, $\psi(x) = 1/(1+x^2)$, which corresponds to the first-order Pad\'e approximation of the Gaussian potential. For the sake of simplicity, we will also restrict our attention to convergent lenses, i.e. $\kappa > 0$. 

The Kirchhoff-Fresnel integral is equivalent to a path integral through the lens plane with phase $S(x, y; \nu, \kappa) = \nu [\frac{(x-y)^2}{2} + \kappa \psi(x)]$. Geometric optics is obtained in the high-frequency limit ($\nu \to \infty$), through the stationary-phase approximation. That is, in geometric optics, the contributions to the Kirchhoff-Fresnel integral reduce to a discrete set of isolated points, or images, $\{ x_j \}$, which have stationary phase, $\partial_x S(x_j) = 0$. The condition of stationary phase gives rise to the lens map, 
\begin{equation}
    \xi(x) \equiv y = x + \kappa \psi'(x),
    \label{eq:lens_map}
\end{equation}
which maps points on the lens plane to a source position $y$. For a given source position, the images, $\{ x_j \}$, with stationary phase, are solutions to the lens equation, $\xi(x) = y$. Note that while the map $\xi$ is one-to-one, the inverse map $\xi^{-1}$ need not be, and a given source position may correspond to multiple images. 

In the geometric, or stationary-phase approximation, each individual image has an associated wave-field given by
\begin{equation}
    F^\mathrm{geom.}_j = \frac{1}{|\Delta_j|^{1/2}} e^{i S(x_j, y; \nu, \kappa) - i \frac{\pi}{2} n_j},
    \label{eq:eik_real}
\end{equation}
where $\Delta_j = \xi'(x_j)$ is the Jacobian of the lens map, and $n_j$ is the Morse index, which is $0$ or $1$ when $x_j$ is a maximum or minimum of the lens map, respectively. The total wave-field at the observer in this limit is given by $F^\mathrm{geom.} = \sum_j F^\mathrm{geom.}_j$ \citep{Schneider}. This is also sometimes referred to as the Eikonal limit of wave optics\footnote{Some authors distinguish between Eikonal and geometric optics, using the term ``geometric optics" to refer to incoherent lensing when there are no interference effects, i.e. when the intensity is given by the incoherent sum $|F^\mathrm{geom.}|^2 = \sum_j |F^\mathrm{geom.}_j|^2$. However, since we only consider coherent lensing described by Eq.~\ref{eq:Kirchhoff-Fresnel} in this paper, we use ``stationary-phase approximation", ``Eikonal limit", and ``geometric optics" interchangeably.}. This approximation is valid in the limit $\nu \to \infty$, except at caustics, which are points where the lens map is degenerate, and $\Delta_j = 0$. At caustics, the intensity of the degenerate images formally goes to infinity. Despite this divergence, geometric optics is an extremely useful approximation in the high-frequency limit as the regions around the caustics at which the divergence occurs typically contribute very little overall flux.

So far we have assumed the lens map is a real map. However, generically the lens potential, $\psi$, and the associated lens map may be analytically continued so that the lens map can admit complex solutions (i.e. the images, $x_j$, may be complex). For such images, the wave-field is given by \citep{GrilloCordes2018}
\begin{equation}
    F^\mathrm{geom.}_j = \frac{1}{|\Delta_j|^{1/2}} e^{i [S(x_j, y; \nu, \kappa) -  \frac{\mathrm{arg}(\Delta_j)}{2}]}.
    \label{eq:eik_imag}
\end{equation}
Such an analytic continuation of the lens map actually improves the geometric approximation, and, indeed, is necessary to capture the behaviour of lensing just outside of caustics \citep{Wright_1980, GrilloCordes2018, Jow2021}. Note, however, that not all complex solutions to the lens equation will contribute to the total flux. In order to describe which complex images are relevant, we will need to briefly introduce the main points of the Picard-Lefschetz theory evaluation of Eq.~\eqref{eq:Kirchhoff-Fresnel} that we discuss in more detail in Appendix~\ref{sec:appendixB}.

At its base, Picard-Lefschetz theory is an application of Cauchy's integral theorem, allowing us to choose a more convenient contour through the complex plane to evaluate Eq.~\eqref{eq:Kirchhoff-Fresnel} over. Picard-Lefschetz theory introduces a set of contours called the Lefschetz thimbles, $\mathscr{J}_j$, each associated with the complex solutions to the lens equation, $x_j$. The thimble, $\mathscr{J}_j$, is defined as the contour of steepest descent in the function $h(x) = \mathrm{Re}\{i S(x, y; \nu, \kappa)\}$ originating from the point, $x_j$. The Cauchy-Riemann equations for an analytic function guarantee that the complex critical points $x_j$ are saddle points, and each $\mathscr{J}_j$ has a corresponding contour of steepest ascent $\mathscr{U}_j$. As we discuss further in Appendix~\ref{sec:appendixB}, the integrand of Eq.~\eqref{eq:Kirchhoff-Fresnel} evaluated along the Lefschetz thimbles is non-oscillatory and exponentially suppressed away from the critical points. The total integral is given by
\begin{equation}
    F(y) = \sqrt{\frac{\nu}{2\pi i}} \sum_{j} N_j \int_{\mathscr{J}_j} e^{i\nu [\frac{(x-y)^2}{2} + \kappa \psi(x)]} dx,
    \label{eq:PLkirchoff}
\end{equation}
where the sum is taken over solutions to the analytically continued lens equation, $x_j$, and $N_j$ is the intersection number between $\mathscr{U}_j$ and the original integration domain (i.e., the real plane). We can see immediately from this how the Eikonal limit arises in the high-frequency limit. The integrand is exponentially suppressed away from the critical points along the Lefschetz thimbles, with a width proportional to $\nu^{-1/2}$. Thus, as $\nu \to \infty$, the contribution to the total integral is confined to an infinitesimally small region around each relevant critical point, $x_j$, recovering the geometric optics result. Now, the relevance of a critical point is determined by $N_j$. If $N_j = 0$, then the critical point does not contribute to the integral and is considered irrelevant. Real critical points, i.e. the real images in geometric optics, are always relevant because the steepest ascent curves, $\mathscr{U}_j$, emanating from them always intersect the real plane. 

Complex critical points with a non-zero imaginary part may or may not be relevant. This is the origin of the claim that not all complex solutions to the lens equation contribute to the geometric/Eikonal approximation in Eq.~\eqref{eq:eik_imag}. Determining which complex images contribute will generically require computing the intersection number $N_j$. While many authors employ a version of geometric optics that neglects complex images, here we will consider the geometric or Eikonal limit to be the fully-general, complex version given by Eq.~\eqref{eq:eik_imag}. One may reasonably ask whether complex images have any physical meaning, or if they are simply a mathematical tool. First, we note that neglecting complex images in the Eikonal limit generically leads to an incorrect value of the field near caustics. In fact, it is possible for the complex images to dominate the observed flux. Secondly, we note that it is possible to compute physical observables associated with complex images, including time delays and position on the sky. Therefore, in all respects, complex images behave physically like real images; see, e.g., \citet{Jow2021} for a more detailed discussion of complex images in lensing.

Now, as we have stated the geometric or Eikonal limit is the high-frequency limit of the Kirchhoff-Fresnel integral. In order to apply this limit, it is necessary to know at what frequency it becomes a good approximation of the diffraction integral. It has sometimes been assumed that this frequency is set by the Fresnel scale \citep[see e.g.][]{Fiedler1987, 1993Natur.366..320C, 2012MNRAS.421L.132P, 2016ApJ...817..176T, dong_extreme_2018, 2022MNRAS.509.5872E},
\begin{equation}
    y_\mathrm{Frnl.} \equiv 1/\sqrt{\nu}.
    \label{eq:Fresnel}
\end{equation}
That is, if $y_\mathrm{Frnl.} \ll 1$, then one is assumed to be firmly in the geometric regime. Note that $y_\mathrm{Frnl.}$ is the dimensionless version of the Fresnel scale given our parametrization of the Kirchhoff-Fresnel integral. Typically, the Fresnel scale is given as a distance scale, $R_F = \sqrt{\lambda \overline{d} / 2 \pi}$, where $\lambda$ is the wavelength of the light. In these units, the relevant limit is when the Fresnel scale is much smaller than the physical scale of the lens. One of the purposes of this work is to argue that the assumption that the Fresnel scale sets the frequency above which geometric optics holds is generically false.

Another tool for approximating the Kirchhoff-Fresnel diffraction integral is the perturbative expansion. In this approximation, one expands $\exp\{i \nu \kappa \psi(x)\} = 1 + i \nu \kappa \psi(x) + ...$ in the integrand of the diffraction integral, giving
\begin{equation}
    F^\mathrm{pert.}(y) = 1 + i \kappa \nu \sqrt{\frac{\nu}{2\pi i}} \int e^{i\nu \frac{(x-y)^2}{2}} \psi(x) dx + \mathscr{O}((\kappa \nu)^2).
    \label{eq:perturb_gen}
\end{equation}
The small parameter that is being expanded in is given by
\begin{equation}
    \epsilon \equiv \kappa \nu
    \label{eq:epsilon},
\end{equation}
which we define here as it will later become of central importance.

The utility of performing this expansion is that for certain potentials, the first term is easier to compute analytically or numerically. For example, for the Gaussian lens, $\psi(x) = \exp\{-x^2/2\}$, the perturbative expansion can be computed to first order analytically as
\begin{equation}
    F^\mathrm{pert.}(y) = 1 + i \epsilon \sqrt{\frac{\nu}{i + \nu}} e^{-\frac{\nu y^2}{2(i+\nu)}}\,
    \label{eq:pert_gauss}.
\end{equation}
Similarly, for the rational lens, $\psi(x) = 1/(1+x^2)$, one can compute the first-order expansion analytically as
\begin{align}
    \nonumber
    F^\mathrm{pert.}(y) & = 1 +i \frac{\epsilon}{2} \sqrt{\frac{-i \nu \pi}{2}} \Big[e^{i \frac{\nu}{2} (y + i)^2} \big(1 + \mathrm{erf}\{\sqrt{\frac{i \nu}{2}} (y+i) \} \big) \\
    &  + e^{i \frac{\nu}{2} (y - i)^2} \big(1 + \mathrm{erfc}\{\sqrt{\frac{-i \nu}{2}} (1+iy) \} \big) \Big],  \label{eq:pert_rat}\\
    &\approx 1 + i \epsilon \sqrt{\frac{-i \nu \pi}{2}} e^{-\nu |y|} e^{i \frac{\nu}{2} (y^2-1)},
    \label{eq:pert_rat_approx}
\end{align}
where the last approximation is valid for large $\nu |y|$. The first-order integral can be computed by first taking the integral in Fourier space by taking the Fourier transform of the potential, $\tilde{\psi}(k) = e^{-|k|}$.

 When the perturbative expansion is valid, the flux at the observer can be described as a small oscillatory modulation about an un-lensed image. That is, when $\kappa = 0$ (i.e. when there is no lens), $F(y) \equiv F_0(y) = 1$, exactly. Thus, in the perturbative limit, the flux is described by a small modulation about the un-lensed flux, $F^\mathrm{pert.}(y) = F_0(y) + i\epsilon g(y)$, for some function $g(y)$. Whereas in geometric optics multiple isolated images form and interfere at the observer, the perturbative regime is effectively described by a single background image plus a small linear modulation. 

\begin{figure}
    \centering
    \includegraphics[width=\columnwidth]{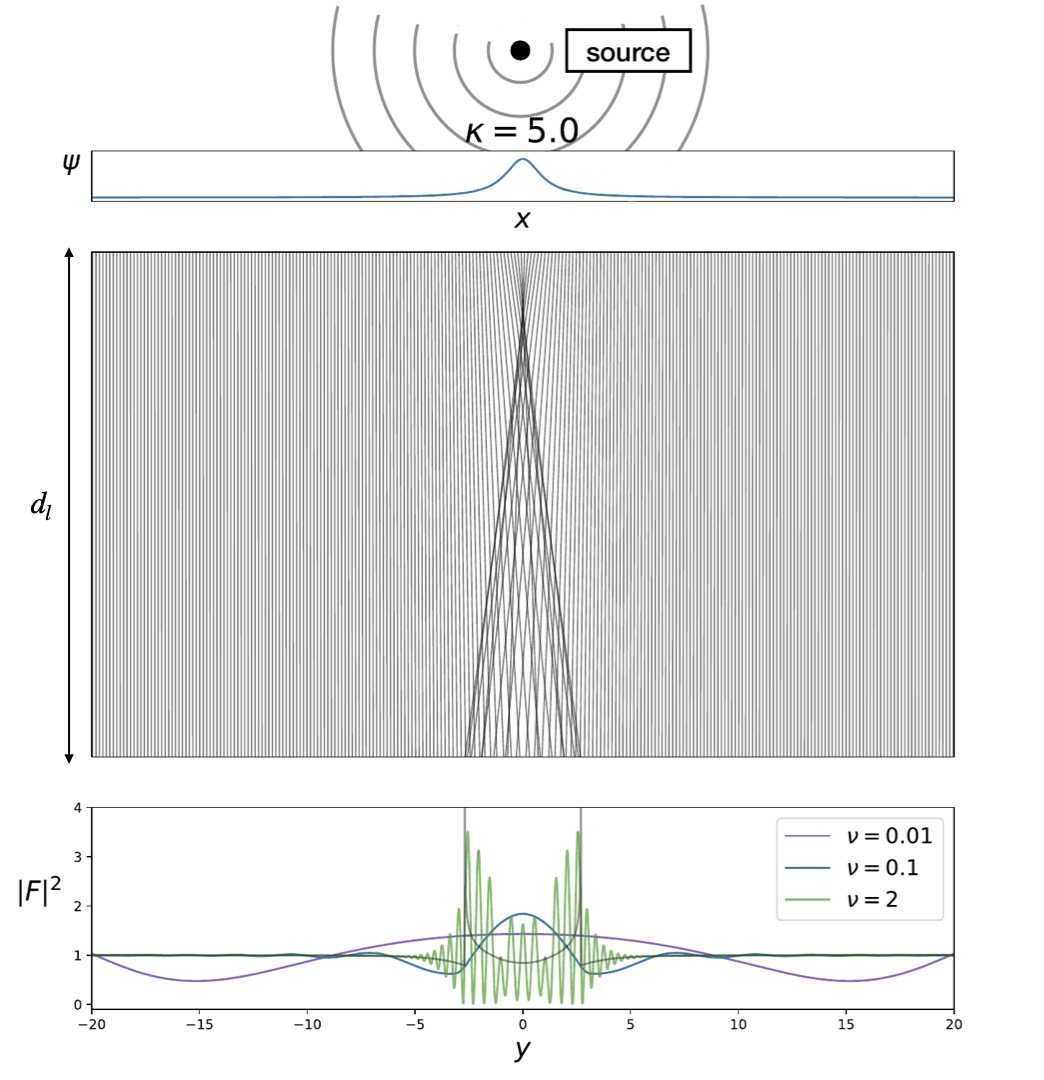}
    \caption{Schematic of thin-plane lensing. The source emits waves that scatter off of the lens potential, $\psi(x)$, localized to a lens plane (top panel). The lens induces a phase shift in the wave-front, which then propagates to the observer. The middle panel shows the light rays propagating from lens to observer, which correspond to the paths of stationary phase connecting source to observer, and are always perpendicular to the wave-front. The observed intensity as a function of observer position is shown in the bottom panel for fixed $\kappa = 5$ and different values of $\nu$ for the lens potential $\psi(x) = 1/(1+x^2)$. Note that the intensity fluctuations for $\nu = 0.01$ are multiplied by a factor of ten so that all three curves can be visualized on the same axis. The grey curve corresponds to the incoherent sum ($\sum_j |F_j|^2$) of the intensities of the images in the geometric limit.}
    \label{fig:ray_diag}
\end{figure}

Fig.~\ref{fig:ray_diag} shows a schematic of the kind of lensing we are considering. A source emits radiation at some distance far from the lens and observer. The waves from the source encounter some lensing potential, $\psi(x)$, localized to the lens plane. This induces a phase variation in the wave front, which then propagates un-interrupted towards the observer, forming an interference pattern in the observer plane. The bottom panel shows the observed intensity computed for the rational lens for a fixed $\kappa = 5$ and different values of $\nu$. For $\nu = 0.01$, the lensing is firmly in the perturbative limit and the intensity is well-described by a small oscillatory modulation about unity (note that in Fig.~\ref{fig:ray_diag} the intensity fluctuations for $\nu = 0.01$ are multiplied by a factor of ten so that all three curves can be plotted on the same axis). For $\nu = 2$, the lensing is firmly in the geometric regime and is described by the interference of multiple images. The curve for $\nu = 0.1$ is in a regime where neither the perturbative nor Eikonal/geometric approximations are good approximations.  

The middle panel of Fig.~\ref{fig:ray_diag} shows the light rays that pass through a given observer position, $y$. These rays correspond to the geometric images that an observer at $y$ observes. Multiple rays may pass through an observer, and regions in the observer plane with different numbers of images per source position are separated by caustics (the caustics are easily visualized in the middle panel of Fig.~\ref{fig:ray_diag}). Now the field at the observer plane can be computed by assuming that each ray corresponds to a point source at the lens plane emitting a spherical wave with a relative phase determined by the lens. The field at the observer is then a sum of all of these spherical waves (this is known as the Huygens-Fresnel principle). Geometric optics arises when the frequencies are large, and so all the waves associated with rays that do not pass through the observer position $y$ cancel out. Thus, the observed intensity in geometric optics is completely determined by the rays passing through a neighbourhood of the observer. In particular, the intensity of the individual images is given by the density of the rays passing through a small neighbourhood of the observer. Thus, in the geometric regime, the intensity curve traces the density of the rays passing through the observer plane. As $\nu$ decreases, the cross section of rays that contribute to the observed intensity at a given source position $y$ becomes finite. As such, the interference between multiple images that is easily visible in the geometric regime gets washed out. 

Note, that the perturbative expansion, which we will later argue can be identified with diffractive optics, explicitly converges when the quantity $\epsilon \ll 1$. This is our first hint that the Fresnel scale may not be the relevant quantity for assessing when geometric optics holds. Given how qualitatively distinct the perturbative (diffractive) and geometric (refractive) descriptions are, the expectation is that there would not be a large region of parameter space where the two regimes overlap. However, if, indeed, geometric optics is valid whenever $y_\mathrm{Frnl.} = 1/\sqrt{\nu} \ll 1$ (or, equivalently, when $\sqrt{\nu} \gg 1$), and the perturbative / diffractive regime occurs whenever $\epsilon = \kappa \nu \ll 1$, then these two regimes should overlap for any given $\nu$ greater than unity, so long as $\kappa$ is chosen to be sufficiently small. Another consequence of this would be that there would be large regions of parameter space that could neither be described by geometric optics, nor by the perturbative expansion. As such, some alternative description of optics in this regime would be called for. 

We argue that this is not, in fact, the case. Rather, contrary to previous assumptions, the Fresnel scale alone is not the appropriate quantity for determining when geometric optics is valid. Instead, a combination of the Fresnel scale and the convergence, $\epsilon = \kappa \nu$, sets both the scale below which the perturbative expansion is valid ($\epsilon \ll 1$), as well as the scale above which geometric optics holds ($\epsilon \gg 1$). In the next section, we will study a specific lens model to make this argument more concrete. In particular, by using Picard-Lefschetz theory to exactly evaluate the Kirchhoff-Fresnel integral at all frequencies, we can quantitatively determine when each of the two approximations we have described are valid.

\section{A simple rational lens}
\label{sec:ratlens}

We will study the simple rational lens with potential given by
\begin{equation}
    \psi(x) = \frac{1}{1+x^2}.
    \label{eq:rat_pot}
\end{equation}
The lens map is given by
\begin{align}
    &\xi(x) \equiv y = x - \frac{2 \kappa x}{(1+x^2)^2}, \label{eq:rat_lensmap}
\end{align}
and its Jacobian takes the form
\begin{align}
    &\xi'(x) = 1 + \frac{2\kappa (3x^2 -1)}{(1+x^2)^3}.
    \label{eq:rat_jacob}
\end{align}
The images corresponding to a position of the observer $y$ -- defined by the identity $\xi(x)=y$ -- are roots of the fifth-order polynomial
\begin{equation}
    x^5 - y x^4 + 2 x^3 - 2 y x^2 + (1-2\kappa) x - y = 0\,.
    \label{eq:rat_poly}
\end{equation}
The quintic polynomial has five roots, either one or three of which being real depending on the position $y$. The real images are always relevant to the integral. The complex images emerge as conjugate pairs (since the polynomial is real), of which one is potentially relevant. By evaluating the intersections of the steepest ascent contour of the complex images and the real line, we determine the relevance of each complex saddle point (see Appendix~\ref{sec:appendixB} for a detailed discussion).

The perturbative approximation of the Kirchhoff-Fresnel integral is given by Eq.~\eqref{eq:pert_rat}. We have chosen to study the rational lens as it has an analytic potential with a similar form to the more widely studied Gaussian lens; indeed, the potential is the first-order Pad\'e approximation of the Gaussian potential $\psi(x) = \exp{(-x^2/2)}$. Thus, we expect the rational lens to behave similarly to the Gaussian lens. However, since the lens equation for the rational lens reduces to a simple fifth-order polynomial, there is only a small number of potentially relevant images, in contrast with the Gaussian lens, for which a potentially infinite number of images may be relevant, making the evaluation of the Kirchhoff-Fresnel integral much simpler for the rational lens. Since, in reality, neither the Gaussian nor the rational lens is likely to be the exact form of any physical lens, we choose to study the simpler case. Once we have found the relevant images for the rational lens, we can evaluate the geometric or Eikonal approximation of the Kirchhoff-Fresnel integral. Our goal is to compare the regimes of validity of the geometric approximation given by Eq.~\eqref{eq:eik_imag} and the perturbative approximation given by Eq.~\eqref{eq:pert_rat}.

\begin{figure}
    \centering
    \includegraphics[width=\columnwidth]{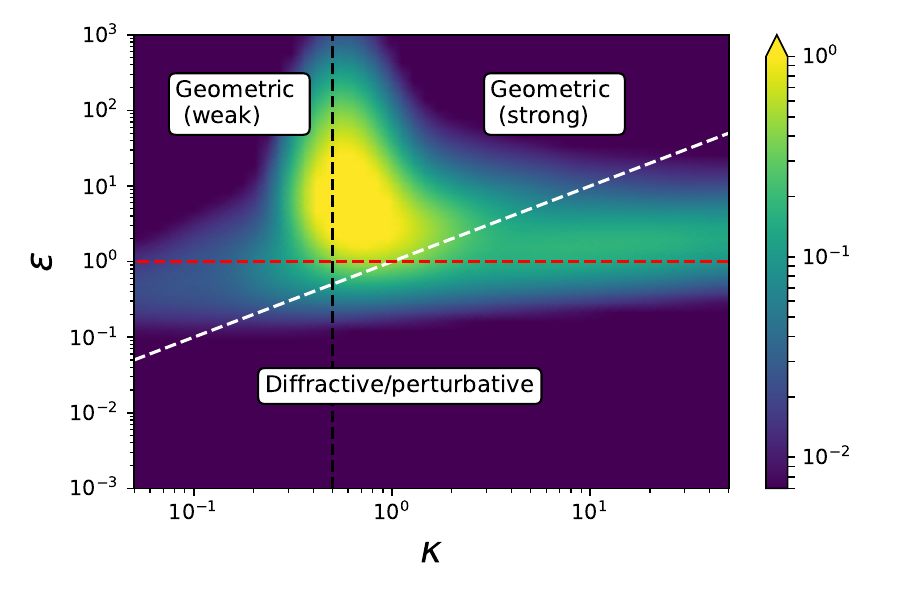}
    \caption{The minimum average residual between the exact value of the intensity of the Kirchhoff-Fresnel integral (computed using Picard-Lefschetz theory) with the perturbative expansion or geometric/Eikonal approximation over a range of source positions, $y$. This residual is defined explicitly in Eq.~\eqref{eq:resid}. For $\kappa > {1/2}$, we choose $L = y_c$, where $y_c$ is the location of the fold caustic. For $\kappa < 1/2$, we choose $L=2$. The residual is shown as a function of $\kappa$ and $\epsilon = \kappa \nu$. The regions are labelled according to the approximation which minimizes the residual. The labels ``strong" and ``weak" refer to the strong and weak lensing regimes discussed in Section~\ref{sec:strong} and corresponding to Table~\ref{table}. The red, white, and black dashed lines correspond to the lines $\epsilon = 1$, $\nu = 1$, and $\kappa = 1/2$, respectively.}
    \label{fig:residual}
\end{figure}

In order to make this comparison, we use Picard-Lefschetz theory to numerically evaluate the Kirchhoff-Fresnel integral. As we discuss in Appendix~\ref{sec:appendixB}, Picard-Lefschetz theory allows for an exact evaluation of the diffraction integral at arbitrary frequencies. Thus, we are able to compare the various approximations to the exact value of the integral. In particular, Fig.~\ref{fig:residual} shows the average minimum residual over a range of source positions, $y$, between the intensity computed by either the perturbative expansion or the geometric/Eikonal approximation with the exact value of the intensity as a function of $\nu$ and $\kappa$. That is, we define the residual to be
\begin{align}
\label{eq:resid}
\text{Res} = \frac{1}{2L} \text{min}&\bigg[\int_{-L}^{L} ||F^{\rm pert.}(y)|^2 - |F(y)|^2| \mathrm{d}y, \\
\nonumber
&\int_{-L}^{L} ||F^{\rm geom.}(y)|^2 - |F(y)|^2| \mathrm{d}y\bigg] / \max_{-L \leq y \leq L}(|F(y)|^2),
\end{align}
where we normalize the residual by the maximum value of the intensity. The exact intensity is computed using Picard-Lefschetz theory, as described in Appendix~\ref{sec:appendixB}. The residual is small when one of the approximations is valid, and becomes large when neither is valid. Note that the geometric/Eikonal approximation diverges for a small region around caustics. To avoid this divergence, we choose $L = \frac{3}{4} y_c$, where $y_c$ is the location of the fold caustic. That is, for the rational lens, when $\kappa > \frac{1}{2}$, the lens exhibits a caustic at $y = \pm y_c$ where $y_c = \frac{3 \sqrt{3}}{8} \kappa$ for $\kappa \gg 1/2$ and $y_c \to 0$ as $\kappa \to 0$. Since most of the flux is contained within the region $-y_c < y < y_c$, we choose $L$ such that we are within this region, in order to avoid the caustics. For $\kappa <= \frac{1}{2}$, we choose $L = 2$. 

Fig.~\ref{fig:residual} shows the residual computed in this way as a function of $\kappa$ and $\epsilon =  \kappa \nu$. From the residual, we can see that both approximations fail at $\epsilon = 1$, which is shown as the red, dashed line. As $\epsilon$ decreases, the perturbative approximation becomes valid, whereas when $\epsilon$ increases, the geometric approximation becomes valid. Thus, the line $\epsilon =1$ forms a boundary between the regimes of validity of these two approximations. This suggests that $\epsilon$ is the appropriate scale that determines when geometric optics holds, as opposed to the Fresnel scale which only depends on the frequency $\nu$. The $\nu = 1$ line is shown as the white, dashed line.  If, indeed, the Fresnel scale were the appropriate separation scale, one would expect the geometric approximation to always hold above this line, and always fail below it. This is clearly not the case, however, as for small $\kappa$, the geometric approximation appears to fail even when $\nu > 1$, and for large $\kappa$ it can remain valid for $\nu < 1$.

It is clear from Fig.~\ref{fig:residual} that the parameter space of the lens is divided into two distinct regimes by the line $\epsilon = 1$. Above this line, geometric optics holds, and below this line, the perturbative expansion holds. As we will argue in Section~\ref{sec:definitions}, geometric optics can be further identified with so-called ``refractive" optics, and non-geometric optics (i.e. the full wave regime) can be identified with ``diffractive" optics. According to this definition, we therefore identify the perturbative regime ($\epsilon \ll 1$) with diffractive optics. There is a regime around $\epsilon \sim 1$ where neither geometric (refractive) optics nor the perturbative (diffractive) expansion holds; however, somewhat surprisingly, the extent of the failure of these approximations is small, and is limited to a rather narrow region about $\epsilon \sim 1$. Therefore, the lensing is almost always well described by either geometric optics or the perturbative expansion. Very rarely is it necessary to compute the full wave integral with more sophisticated techniques.

Note that Fig.~\ref{fig:residual} also highlights the boundary at $\kappa = 1/2$, which is where the cusp caustic occurs. The cusp occurs when the two fold caustics merge, and $y_c \to 0$. The increase in the residual around this line is due to the fact that as $y_c \to 0$, our selection of the range $-y_c < y < y_c$ no longer avoids the caustics. However, as $\epsilon$ increases, the region around the caustic for which the geometric/Eikonal approximation diverges becomes smaller. We label the two regions to the left and right of $\kappa = 1/2$ weak and strong geometric optics, respectively, as it is only when $\kappa > 1/2$ that multiple real images are formed by the lens, as we discuss further in Section.~\ref{sec:strong}. 
 
\begin{figure*}
    \centering
    \includegraphics[width=2\columnwidth]{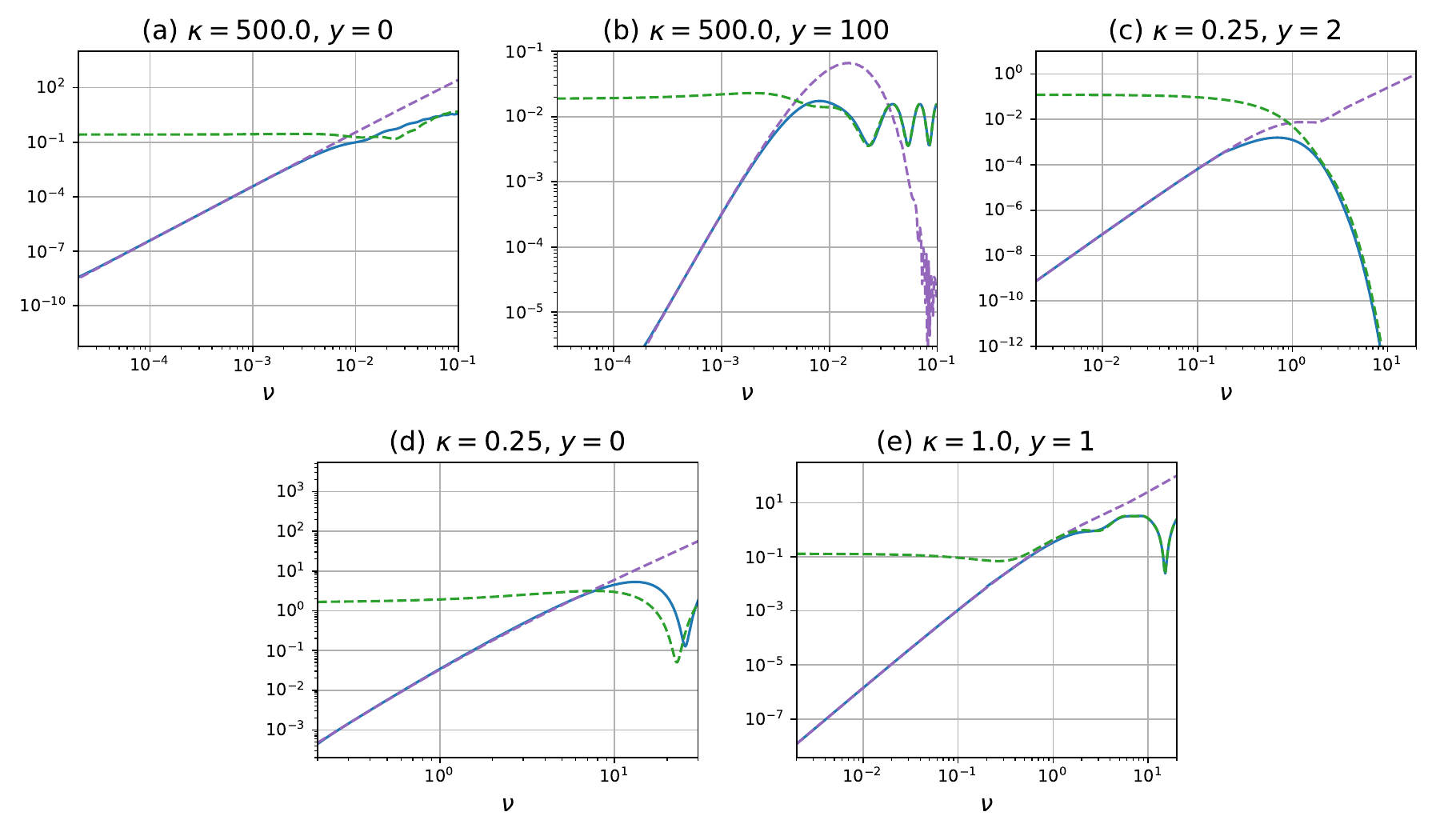}
    \caption{Comparison of the absolute value of the wave-field minus one, $|F(y) - 1|^2$, as a function of frequency for the perturbative/diffractive expansion (dashed purple lines), geometric optics (dashed green lines), and the exact evaluation of the Kirchhoff-Fresnel integral computed using Picard-Lefschetz theory (blue). The curves are computed for five different values of $\kappa$ and $y$, which correspond to values in the five topologically distinct regions of the rational lens shown in Fig.~\ref{fig:topology}.}
    \label{fig:transition}
\end{figure*}

Fig.~\ref{fig:transition} further examines the transition from the geometric regime to the diffractive/perturbative regime by comparing the asymptotic behaviour of the exact Kirchhoff-Fresnel integral, the perturbative expansion, and the geometric/Eikonal approximation as a function of frequency for specific values of $\kappa$ and $y$. Five values of $\kappa$ and $y$ were chosen, corresponding to the five distinct regions discussed in Appendix~\ref{sec:appendixB}. In short, these five regions correspond to the five distinct Lefschetz thimble topologies for this lens, or, in other words, each region has a unique number of real and relevant complex images. What is shown in Fig.~\ref{fig:transition} is not the total intensity, $|F(y)|^2$, but rather the absolute value squared of the total wave-field minus one, $|F(y) - 1|^2$. This way we are comparing the value of the asymptotic scaling of the perturbative part in the diffractive regime. The exact value, computed using Picard-Lefschetz theory, is shown in blue, whereas the perturbative expansion and geometric optics results are shown in purple and green, respectively. 

As expected, regardless of the choice of source position and lens strength, the geometric limit returns the correct asymptotic behaviour for large frequencies (except at caustics), whereas the perturbative/diffractive limit assumes the correct asymptotic behaviour for low frequencies. What is notable, however, is that the frequency where this transition occurs depends on the lens strength $\kappa$, and, in particular, is given to an order-of-magnitude by $\epsilon=1$. In the region where both approximations agree with each other (around $\epsilon = 1$), neither approximation agrees with the true result, but unlike with small frequencies, where geometric optics is off by orders-of-magnitude, and large frequencies, where the perturbative expansion is off by orders-of-magnitude, both are only off by a few factors of unity. It is particularly striking that for large lens strengths, the geometric approximation can hold for very low frequencies (for example, when $\kappa = 500$ and $y=100$, geometric optics holds down to $\nu \sim 10^{-2}$). Similarly, when $\kappa$ is small, geometric optics can be made to fail for large frequencies (when $\kappa = 0.25$ and $y=0$, the geometric approximation fails as early as $\nu \sim 10$). 

We infer two things from these results. First, that, together, geometric optics and the perturbative expansion can be used to accurately describe most of parameter space, except for a small region where they overlap. Secondly, the quantity that determines the separation between these two regimes is $\epsilon = \kappa \nu$, as opposed to the Fresnel scale. Expressed in terms of dimensionful quantities, $\epsilon = \kappa a^2 / R_F^2$, where $a$ is the size of the lens (see Section~\ref{sec:turbulence}). Thus, the physical scale that separates the diffractive and refractive regimes is given by $R_F / \sqrt{\kappa}$ (i.e. when the size of the lens is much larger than this scale, refractive optics holds, and when it is much smaller, diffractive optics holds). This means that the convergence plays an important role in determining whether geometric optics holds, rather than the frequency alone. Thus, the simple heuristic that geometric optics is valid when the wavelength is small compared to the size of the lens is generically false.

\section{Refractive vs. diffractive optics}
\label{sec:definitions}

The primary concern of this paper is to delineate regimes of optics in the context of lensing. As such, it is important to make clear the definitions of the words we use to describe these regimes. Words such as ``diffraction", ``refraction", and ``interference" are used widely in the astronomy and physics literature, often suggesting different meanings when used by different authors. In this section, we will attempt to clearly explain our definitions of these words, why we have chosen these definitions, and how they differ from other authors' definitions. This will be particularly important for our discussion of interstellar scintillation later on.

To begin with, we note that we are only considering the lensing of coherent sources, as described by the Kirchhoff-Fresnel integral for a monochromatic plane wave. Therefore, all effects we are considering are ``interference" effects. The primary observable we are interested in calculating is an ``interference pattern". This is the case even when the Kirchhoff-Fresnel integral is well-described by the stationary-phase approximation, which we refer to as Eikonal or geometric optics. Note also that some authors will use ``diffraction" to refer to any kind of coherent interference effect. Here, we distinguish between ``interference" and ``diffraction", with the former being any kind of coherent wave effect, and the latter referring to a particular regime of coherent lensing.

We will define the terms ``refractive optics" and ``diffractive optics" in the following way:
\begin{itemize}
    \item \textbf{Refractive optics} = Geometric optics. In this limit, the Kirchhoff-Fresnel integral is well-described by the stationary phase approximation. The observed wave-field is a coherent sum of discrete images. The images are formed by rays that are bent by the lens into the line of sight of the observer, with a bending angle given by the index of refraction of the lens according to Snell's law. 
    \item \textbf{Diffractive optics} = Non-geometric optics. This regime is characterized by the breakdown of the stationary phase approximation. The simple image picture of geometric optics is no longer valid, and one must perform the full wave calculation to determine the observed interference effects. 
\end{itemize}
Our numerical results shown in Fig.~\ref{fig:residual} and discussed in Sec.~\ref{sec:ratlens} further motivate this definition. Our results show that lensing can be well separated into two distinct regimes characterized by the validity of geometric optics ($\epsilon > 1$) and the validity of the perturbative expansion of the Kirchhoff-Fresnel integral ($\epsilon < 1$). It is sensible to identify refractive optics with the former and diffractive optics with the latter.

Our simple, mathematically motivated definitions of these terms is in contrast with observationally motivated definitions that have arisen in the pulsar scintillation literature. In that literature, diffractive interstellar scintillation (DISS) and refractive interstellar scintillation (RISS) refer to observational phenomena \citep{Rickett1990}. In particular, one finds that for a turbulent ISM, the power in the intensity fluctuations of scintillation as a function of spatial scale follow distinct scaling relations above and below the Fresnel scale. While the particular scaling relationships will depend on the details of the power spectrum of the turbulent medium, in general below the inverse Fresnel scale, $R_F^{-1}$, the power increases with wave-number, $k$, according to some power law, and decreases according to a different power-law in $k$ above $R_F^{-1}$ \citep[see Fig.~1 in][]{Rickett1990}. It is this distinct behaviour in the observed power spectrum of intensity fluctuations that define DISS and RISS, with the former referring to the small-scale behaviour and the latter to the large-scale behaviour.

It has often been assumed that RISS is described by large-scale inhomogeneities in the ISM forming multiple images by bending rays into the line of sight according to the index of refraction, and DISS is described by small-scale inhomogeneities which induce higher-order wave effects (indeed, this is why these observationally defined regimes were given the names ``refractive" and ``diffractive" to begin with). See, for example, \citet{1986ApJ...310..737C} which distinguishes the physics of RISS and DISS according to the stationary-phase approximation, which is in-line with our definition of ``refractive" and ``diffractive" optics. More fundamentally, it is assumed that DISS is produced by light coming from a typical bending angle given by $\alpha^* = \lambda / a$, where $\lambda$ is the wavelength of light and $a$ is the physical size of the plasma inhomogeneity in the ISM, whereas RISS may arise from fluctuations on much larger angular scales. As we will show, our definition of refractive and diffractive optics also aligns with this assumption.

Why then are we making a distinction between the mathematically motivated definition of refractive and diffractive optics and the observational phenomena of DISS and RISS? There are a few reasons for this. Firstly, identifying RISS and DISS with refractive and diffractive optics (as we have defined them) relies on the description of the ISM as a turbulent medium with a simple power-law spectrum between some inner and outer scales. Yet, in the decades since the discovery of parabolic arcs in scintillation secondary spectra \citep{2001ApJ...549L..97S} it has been suggested that purely refractive models can describe all of the observational phenomena, including what has been labelled DISS \citep{2006ApJ...640L.159G, 2014MNRAS.442.3338P}. Thus, we need to be careful not to conflate the physical regime of optics with the form of the ISM.

Secondly, DISS is, by definition, a strong phenomenon. That is, DISS refers to a regime where the modulation index of scintillation is large (i.e. the observed intensity fluctuations are of order unity). Mathematically, this means that the induced phase fluctuations in the wavefront of the propagating light must be large. However, as we have seen, large phase variations ($\epsilon \gg 1$) necessarily imply the validity of the stationary-phase approximation, i.e. refractive optics. That is, we will argue that diffractive optics is a necessarily weak phenomenon. This is more than a mere semantic argument. Crucially, if correct, this would mean that the standard assumption that the characteristic scaling relationship of intensity fluctuations associated with DISS automatically implies the angular relationship $\alpha^* = \lambda / a$ is faulty. That is, standard inferences of the physical size of the plasma structures responsible for lensing in the ISM may be incorrect.

\section{Strong scintillation in the refractive and diffractive regimes}
\label{sec:strong}

The results of Sec.~\ref{sec:ratlens} suggest that it cannot always be taken for granted that short wavelengths automatically mean geometric optics holds. Rather, one must also know the lens strength, or convergence, in order to assess whether one is in the geometric or diffractive regime. At first glance, this seems like a challenge, as the convergence is a much more difficult quantity to measure. However, in this section, we will make a simple scaling argument that an observation of large flux variations implies that the lensing is taking place in the geometric regime. In other words, diffractive lensing is typically weak. Thus, for many practical purposes (e.g. pulsar scintillation and extreme-scattering events), any lensing observations are likely to occur in the geometric/refractive optics regime.

For an isolated lens, such as the rational lens we have been considering, it is straightforward to show that strong lensing can only occur in the geometric regime. In the perturbative regime ($\epsilon \ll 1$), the peak intensity occurs when $y=0$; in particular, $|F^\mathrm{pert.}(0) - 1|^2 \sim \epsilon^2 \nu = \kappa^2 \nu^3$. In order for this to be of order unity or larger and for the perturbative expansion to hold, we require $\kappa^2 \nu^3 \gtrsim 1$ and $\kappa \nu \ll 1$, which together imply that $\kappa^{-2/3} \lesssim \nu \ll \kappa^{-1}$. Thus, the maximum intensity fluctuations are only ever larger than unity in the perturbative regime in a small region of parameter space when the lens strength $\kappa$ is extremely small. By contrast, the peak intensity can be made arbitrarily large in the geometric/refractive regime as $\epsilon \to \infty$. Thus, for an isolated lens, large magnifications imply refraction, and diffraction is a weak phenomenon.

\subsection{A uniformly spaced ensemble of rational lenses}
\label{sec:ensemble}

\begin{figure}
    \centering
    \includegraphics[width=\columnwidth]{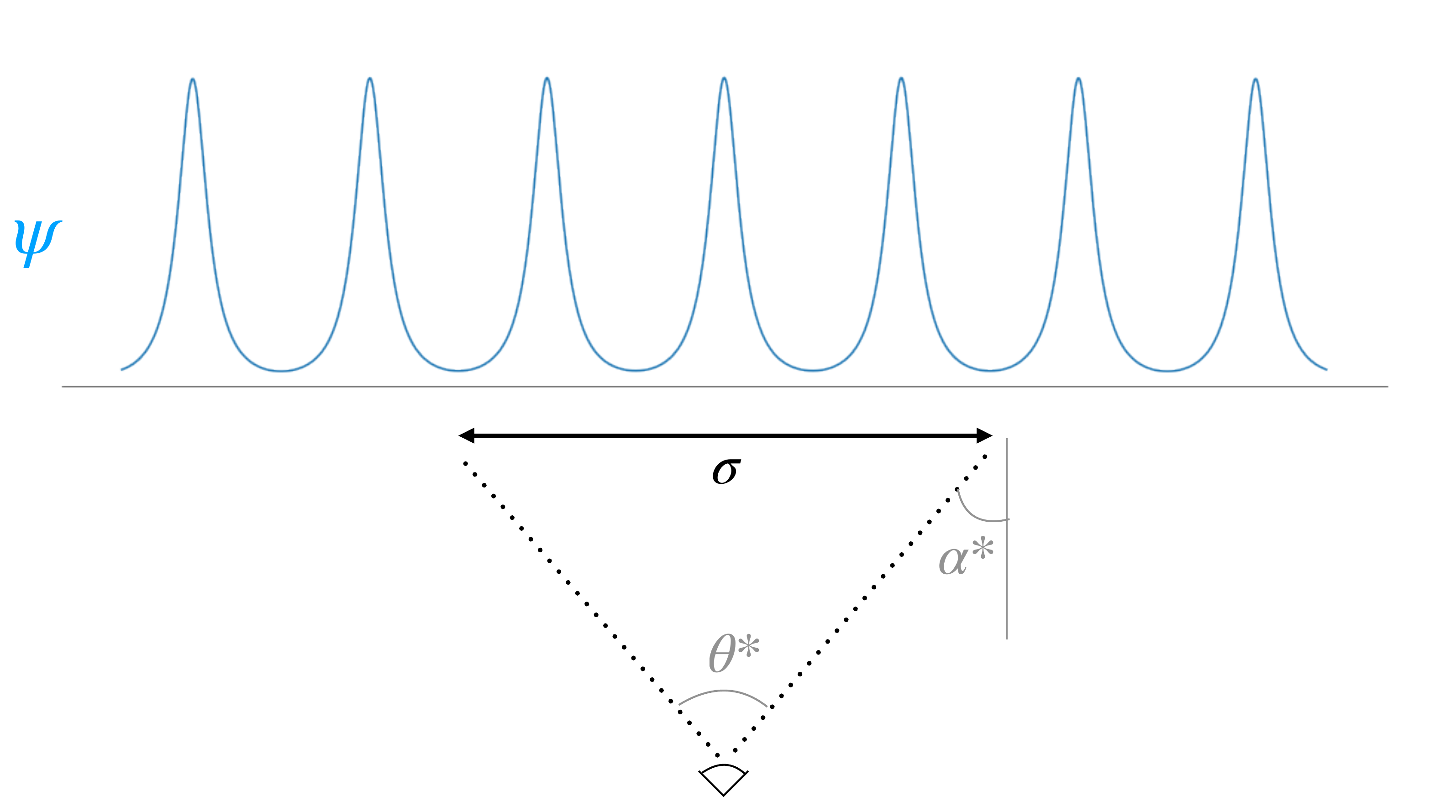}
    \caption{An ensemble of well-separate rational lenses (i.e. the separation between each peak is larger than the width of the peaks). The cross-section $\sigma$ refers to the size of the region in the lens plane within which the individual lenses contribute significantly to the flux, and $\theta^*$ is the angular size of that region.}
    \label{fig:ensemble}
\end{figure}

We have shown that an isolated lens cannot produce large intensity variations in the diffractive regime. However, for the study of scintillation, we are primarily interested in extended plasma lenses with multiple peaks and troughs. Specifically, scintillation has often been studied assuming a turbulent, Gaussian random field model for the plasma structures responsible for scintillation \citep{1975ApJ...196..695L, 1977ARA&A..15..479R}. It is challenging to extend our results for the single isolated lens to the more complicated case of a general Gaussian random field, but we hope to gain some insight into this more general problem by considering the simple case of many rational lenses added together with some fixed spacing as shown in Fig.~\ref{fig:ensemble}. We hope to further this work with a more complete study of the Gaussian random field lens.

 We will consider the case when the spacing between the lenses is large enough so that there is not significant overlap between the lenses. We are interested in determining under what circumstances the resultant scintillation can be described as ``strong" or ``weak". It is possible to define strong versus weak lensing in many ways. Here we will consider the lensing to be strong when the average intensity fluctuations are of order unity or larger, and the lensing to be weak when it is much less than one. This average intensity fluctuation is sometimes referred to as the scintillation index, $m$. To make our argument, we will study an ensemble of identical rational lenses, evenly spaced such that the distance between peaks is given roughly by the width of each lens, $a$, as shown in Fig.~\ref{fig:ensemble}. We will consider four limiting cases: when $\kappa \gg 1$ and $\kappa \ll 1$, with $\epsilon \gg 1$ or $\epsilon \ll 1$.

The basic argument is as follows: for an isolated lens, there is a certain lateral displacement that the lens can undergo before the observed flux falls off exponentially. In other words, there is a value of the impact parameter, $y^*$, for which $|y| > y^*$ implies that the observed flux, $|F(y)|^2$, is exponentially suppressed. This is generally true, regardless of the particular regime of optics. This defines an effective cross-section for a given lens, $\sigma = 2 y^*$. That is, if the lens lies outside of a region of width $\sigma$ relative to the observer, then the lens will effectively not contribute to the observed flux. Thus, only the lenses within the area $\sigma$ in Fig,~\ref{fig:ensemble} contribute to the flux at the observer. If we can compute the average intensity modulations, $\mu$, of an isolated lens over the range $|y| < \sigma / 2$, then the total intensity modulations due to the ensemble of lenses is roughly $m \sim \sigma \mu$. It follows that all we need to do is compute the values of $\sigma$ and $\mu$ for the isolated lens in the different regimes, the results of which are shown in Table~\ref{table}.

\begin{figure}
    \centering
    \includegraphics[width=\columnwidth]{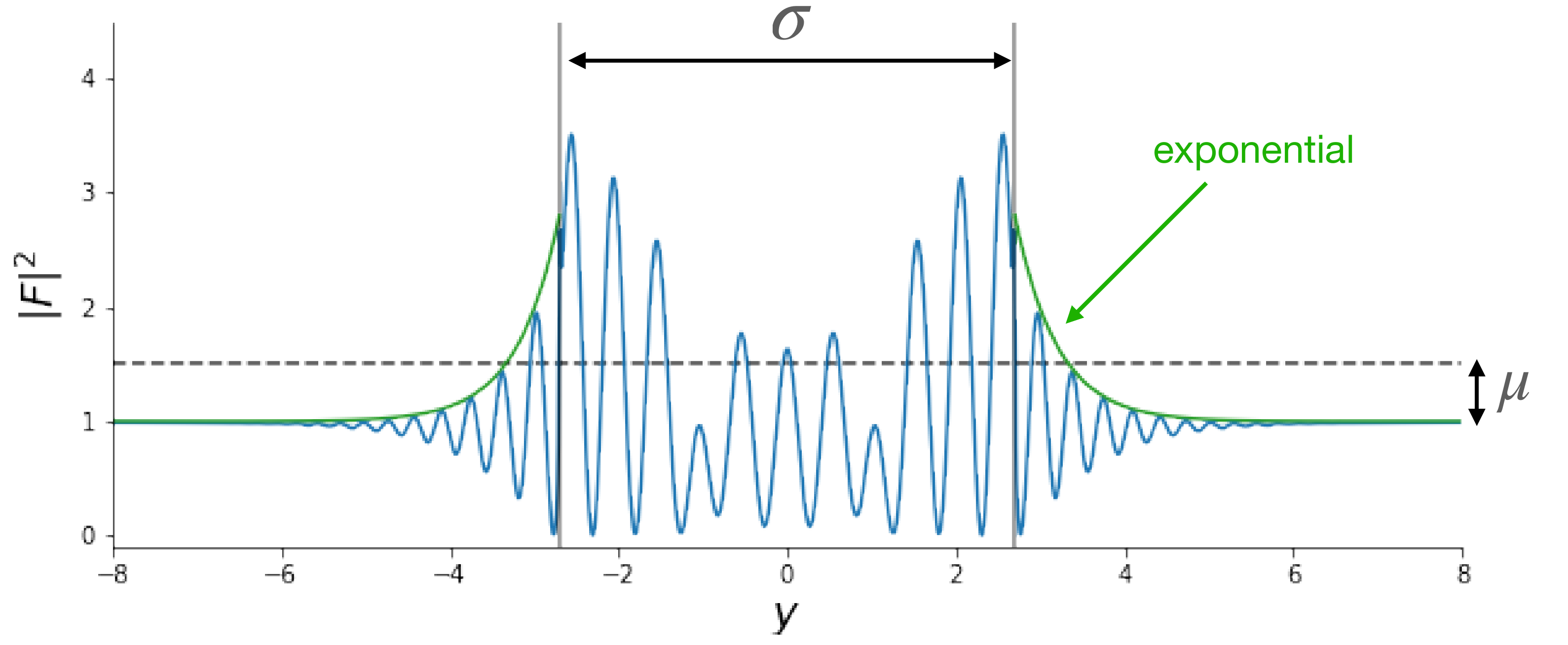}
    \caption{Intensity fluctuations as a function of $y$ for the rational lens with $\kappa = 5$ and $\nu = 2$. Outside of a region of width $\sigma$, the intensity fluctuations decay exponentially. Within the region, the intensity fluctuations have an rms of $\mu$.}
    \label{fig:flux_diag}
\end{figure}

Now as we have shown, when $\epsilon \gg 1$ we are in the geometric regime and when $\epsilon \ll 1$ we are in the diffractive or perturbative regime. In either case, outside of a certain region centred around the origin, the intensity above unity, $|F|^2 - 1$, falls off exponentially. This is easy to see for the perturbative/diffractive regime from Eq.~\eqref{eq:pert_rat_approx}. This is a little more difficult to see for geometric optics, but as we show in Appendix~\ref{sec:appendixB}, for any $\kappa$ there exists a $y$ beyond which the only relevant images are a single real image and one or two imaginary images. The real image is effectively unperturbed and has an intensity of order unity, and the imaginary images have an intensity that has an exponential dependence on the impact parameter that goes like $~e^{-i\nu y}$. Thus, our assumption that the flux from an isolated lens is limited to a region in the observer plane (which we call $\sigma$) is valid. Fig.~\ref{fig:flux_diag} shows an illustrative example computed for $\kappa = 5$ and $\nu = 2$ in the geometric regime. Outside of a region of width $\sigma$, the intensity peaks decay exponentially, and the rms intensity fluctuations within the region are given by $\mu$. The total rms fluctuations of an ensemble of such lenses is then given by $m = \sigma \mu$. The lensing is considered weak when $m \ll 1$ and strong when $m \gtrsim 1$.

First, let us consider $\kappa \gg 1$. If $\epsilon \gg 1$, then we are in the geometric regime. Since $\kappa > 0.5$, the value of $\sigma$ is set by the region between the two caustics (see Fig.~\ref{fig:topology}). For $\kappa \gg 1$, the area between the caustics scales as $\sigma \sim \kappa$. Now, when $\kappa$ is large, using Eq.~\eqref{eq:rat_jacob} and Eq.~\eqref{eq:eik_imag}, the intensity fluctuations, which we will call $\mu$, are of order $\sim \kappa^{-1}$. Thus, the total rms fluctuations are $m = \sigma \mu \sim 1$, which we consider strong lensing. When $\epsilon \ll 1$, we are in the perturbative/diffractive regime. From Eq.~\eqref{eq:pert_rat_approx}, the cross section is given by $\sigma \sim \nu^{-1}$, and the intensity fluctuations are $\mu \sim \epsilon^2 \nu$. Thus, in this regime, $m = \sigma \mu \sim \epsilon^2 \ll 1$. So, when $\kappa$ is large, strong lensing is never described by the perturbative or diffractive regime. 

Now, if $\kappa \ll 1$, there are no caustics. Thus, for $\epsilon \gg 1$ (the geometric regime), there is only one real image, and potentially multiple imaginary images. The imaginary images, however, are exponentially suppressed by a factor of $\sim e^{-\epsilon}$. The intensity of the real image is straight forward to compute since the real image occurs at roughly $x \approx y$. Thus, for small $\kappa$ and large $y$, we have $|\Delta|^{-1} \approx 1 + \frac{6\kappa}{y^4}$. Thus, the intensity fluctuations are small when $y \gg 1$, so we take $\sigma \sim 1$. When $y \lesssim 1$, the intensity fluctuations are of order $\mu \sim \kappa$, and $m = \sigma \mu \sim \kappa \ll 1$. When $\epsilon \ll 1$ (the perturbative regime), we have again that $\sigma \sim \nu^{-1}$ and $\mu \sim \epsilon \nu$, so $\sigma \mu \sim \epsilon$. Thus, when $\kappa \ll 1$, regardless of whether we are in the refractive or diffractive regime, the lensing is weak. Table~\ref{table} summarizes these results. We have argued that strong lensing generally occurs when $\kappa \gg 1$ and $\epsilon \gg 1$, but not when either $\kappa \ll 1$ or $\epsilon \ll 1$. In other words, any lensing described by the perturbative or diffractive regime is weak. 

Here we have used our results for a single, isolated lens to reason about a uniformly spaced ensemble of such lenses. This model, however, is qualitatively very similar to a sinusoidal phase screen (with amplitude given by $\kappa$ and wavelength given by $a$), for which there exists an analytic expression for the Kirchhoff-Fresnel integral \citep{1997RaSc...32..913B}. Comparison with quantitative results for the sinusoidal phase screen shows that, indeed, strong scintillation ($m \gtrsim 1$) can only occur when $\epsilon = \kappa \nu \gg 1$. 

It is also possible to compute the associated angular size of the scattering disk on the sky for the ensemble of lenses we have considered in the different optical regimes. That is, we can convert the effective cross-section $\sigma$ into an angle $\theta^*$, which describes the range of angles over which the majority of the observed flux is coming from. The cross-section $\sigma$ is the value of $y$ over which the lens contributes to the flux. In terms of dimensionful quantities, $y = \hat{y} / a$ where $\hat{y} = (\hat{x}_s d_l + \hat{x}_o d_{sl}) / d_s$. If we take the position of the source relative to the lens to be fixed, then the effective range of angular positions on the sky over which a given isolated lens will contribute to the observed flux is $\theta^* = \sigma a / d_l$. Thus, the perturbative regime generically results in the angular size of the scattering disk to be $\theta^* \sim a\nu^{-1} / d_l \sim (\lambda / a) (d_{sl} / d_s)$. Geometrically, we can convert the angular size of the scattering disk to a bending angle by $2 \alpha^* = \theta^* d_s / d_{sl}$, which gives $\alpha^* \sim \lambda / a$ in the perturbative regime. This is another reason why it is sensible to identify the perturbative regime with diffractive optics, as this characteristic scattering angle that scales linearly with wavelength and inversely with the size of the lens has often been associated with the so-called diffractive interstellar scattering (DISS) phenomenon. This, however, has interesting implications for the study of interstellar scintillation, as we will discuss in Section~\ref{sec:turbulence}. DISS, is by definition, a strong phenomenon; however, our results suggest that a $\alpha^* \sim \lambda/a$ scattering angle is only ever associated with weak scintillation.

\begin{table*}
\begin{center}
\begin{tabular}{c|c|c|c|c|c|c|c}
$\kappa$ & $\epsilon = \kappa \nu$    & regime & cross section ($\sigma$) & intensity fluctuations ($\mu$) & $m = \sigma \mu $ & $\alpha^*$ & strong/weak  \\
\hline
$\gg 1$  & $\gg 1$     & geometric / refractive & $\sim \kappa$ &
$\sim \kappa^{-1}$ & $\sim 1$ & $\kappa a / \overline{d}$ & strong
\\
\hline
$\gg 1$  & $\ll 1$     & perturbative / diffractive & $\sim \nu^{-1}$ &
$\sim \epsilon^2 \nu$ & $\sim \epsilon^2$ & $\lambda / a$ & weak \\
\hline
$\ll 1$  & $\gg 1$     & geometric / refractive & $\sim 1$ &
$\sim \kappa$ & $ \sim \kappa $ & $a / \overline{d}$ & weak    \\
\hline
$\ll 1$  & $\ll 1$     & perturbative / diffractive & $\sim \nu^{-1}$ &
$\sim \epsilon^2 \nu$ & $\sim \epsilon^2$ & $\lambda / a$ & weak  \\
\hline
\end{tabular}
\caption{Summary of the scaling arguments for the strength of lensing in the different regimes for the simple rational lens. Strong lensing never occurs when $\kappa \ll 1$, or when the perturbative expansion holds.}
\label{table}
\end{center}
\end{table*}

\section{Measuring time delays}
\label{sec:tdelays}

In the previous section, we argued that strong scintillation generically occurs in the geometric regime, whereas diffractive optics generically produces weak scintillation, as the total rms intensity fluctuations only becomes of order unity in the geometric regime. However, while the total rms fluctuations may increase, even if the intensity fluctuations are large (especially near caustics), if the resolution of an observation is too small, the actual observed intensity will be an average over many oscillations, and any large intensity fluctuations will be washed out. How then are we to distinguish geometric optics from diffractive optics? Fortunately, the two regimes behave qualitatively very distinctly in delay space. 

Observationally, the quantity one typically observes is the wave-field as a function of frequency and time, $F(\nu, t)$, where the time dependence comes from the relative motion of the source and lens as a function of time, $y(t)$. Here we will focus on the wave-field as a function of frequency. Typically, one actually observes the dynamic spectrum, which is the intensity, $|F|^2$, as a function of frequency and time; however, novel phase retrieval techniques have made observations of the underlying wave-field possible \citep{2021MNRAS.500.1114S}. 

Consider lensing in the geometric regime when multiple images are present. Then the wave-field takes the form
\begin{equation}
    F^\mathrm{geom.}(\nu) = \sum_j \frac{1}{|\Delta_j|^{1/2}} e^{iS(x_j,y;\nu,\kappa) - i \frac{{\rm arg}(\Delta_j)}{2}},
    \label{eq:geom_repeat}
\end{equation}
as in Eq.~\eqref{eq:eik_imag}. Now, consider taking a Fourier transform of the wave-field as a function of frequency. In the geometric regime, the result is
\begin{equation}
    \hat{F}_\nu \propto \sum_j \frac{1}{|\Delta_j|^{1/2}} \delta(\tau - T(x_j, y, \kappa)),
\end{equation}
where $\hat{F}_\nu $ denotes the Fourier transform of $F$ as a function of $\nu$, $\tau$ is the conjugate variable of $\nu$, and $T(x,y,\kappa) \equiv \partial_\nu S = \frac{1}{2}(x-y)^2 + \kappa \psi(x)$ is the dimensionless time-delay associated with an image at $x$. Thus, in the geometric regime, the wave-field in delay space is a sum of delta functions centred at well-defined time delays associated with each image. Conversely, as we argue in Appendix~\ref{sec:appendixB}, in the low-frequency limit, the geometric images become mutually coherent, so that the diffractive/perturbative wave-field, rather than being characterized by an interference pattern of multiple incoherent images, effectively becomes a single image with a small modulation on top. Thus, one would not expect the diffractive/perturbative wave-field to exhibit distinct images in delay space.

\begin{figure}
    \centering
    \includegraphics[width=\columnwidth]{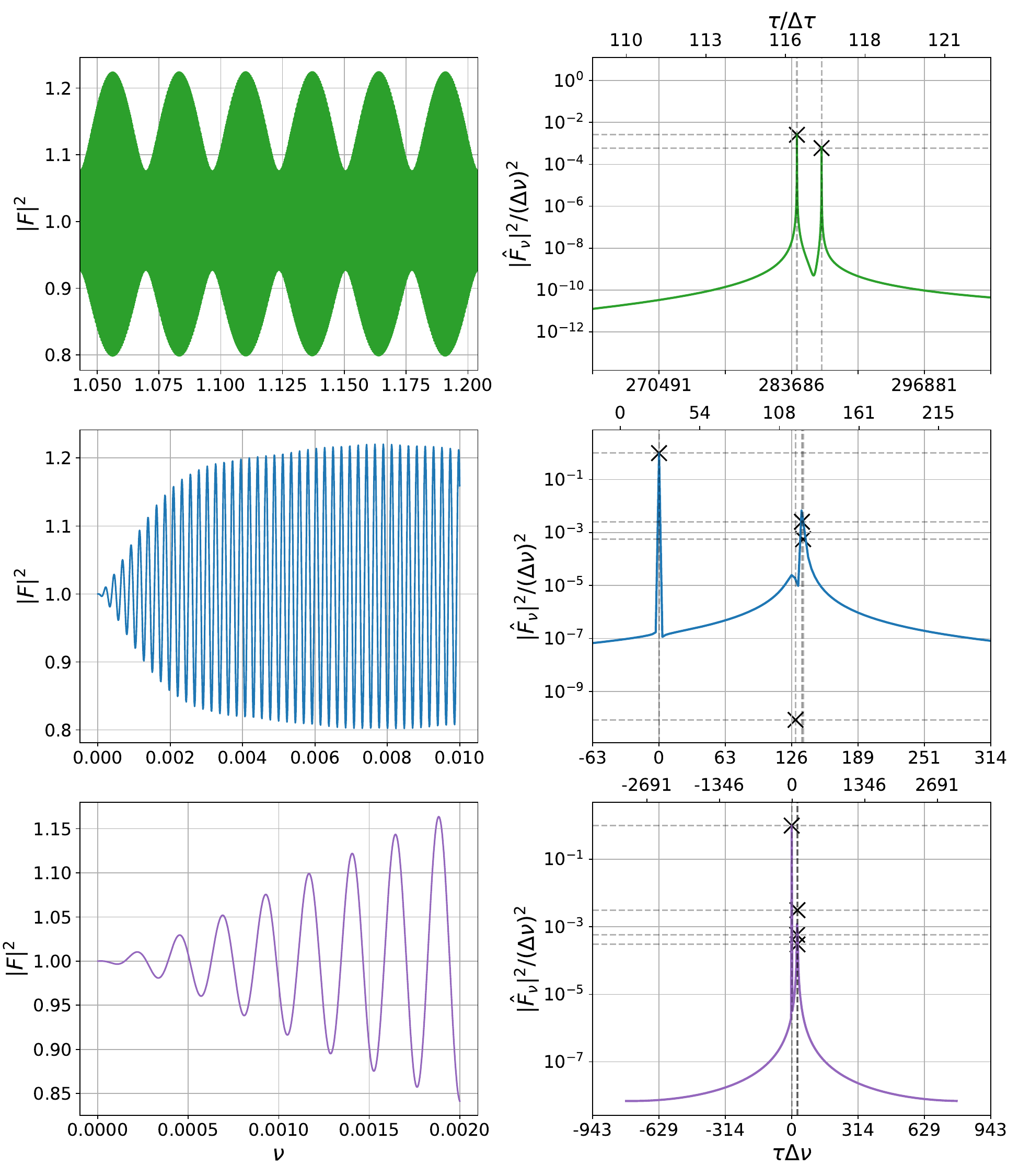}
    \caption{(Left column) The intensity $|F(\nu)|^2$ of the rational lens computed for $\kappa = 500$ and $y=162$, for different ranges of frequency, $\nu$. The frequency ranges are chosen such that the top panel occurs firmly in the geometric regime ($\epsilon = \kappa \nu \gg 1$), the bottom panel occurs in the perturbative regime ($\epsilon < 1$), and the middle panel occurs in an intermediate regime. (Right column) The Fourier transform of the wave-field as a function of $\nu$, denoted by $\hat{F}_\nu (\tau)$, where $\tau$ is the conjugate variable of $\nu$. The vertical axis shows $|\hat{F}_\nu / \Delta \nu|^2$ where $\Delta \nu$ is the frequency bandwidth for which the wave-field was computed. Under this normalization, the conjugate wave-field in the geometric optics regime is a sum of isolated peaks centred at the time delay for each image, with a height given by the individual intensity of the image in the geometric optics approximation. The black X's in the right column correspond to each of the images' individual intensities and time delays. Picard-Lefschetz theory makes it possible to assign individual intensities to the relevant images even at small frequencies. The horizontal axis of the left column is shown in two units: units of the fundamental delay resolution given by $(\Delta \nu)^{-1}$, and units of the relative time delay $\Delta \tau$, which is the distance between the two peaks shown in the top right panel.}
    \label{fig:ft}
\end{figure}

Fig.~\ref{fig:ft} shows the intensity fluctuations and the conjugate wave-field, $\hat{F}_\nu$, for the rational lens with $\kappa = 500$ and $y = 162$, which is halfway to the caustic ($y = y_c/2$). The lensing is computed for three different frequency ranges, $\nu \in [0, 0.002]$, $[0, 0.01]$, and $[0, 20]$, where the ranges are chosen so that the lensing occurs firmly in the diffractive/perturbative regime for the smallest range, firmly in the geometric/Eikonal regime for the largest range, and neither for the intermediate range.

For the choice of $\kappa$ and $y$, there are four relevant images: three real, and one imaginary. In other words, the sum in Eq.~\eqref{eq:geom_repeat} contains four terms, where $x_j$ is purely real for three of those terms. As a result, the conjugate wave-field for the geometric regime (the top panel) is, in principle, a sum of four delta functions. Due to the finite bandwidth, $\Delta \nu$, for which we calculate the wave-field, the conjugate wave-field for the geometric regime is actually given by $|F_\nu|^2 \sim \sum_j (\tau - T(x_j))^{-2}$. The right column of Fig.~\ref{fig:ft} shows the amplitude of the conjugate wave-field normalized by the frequency bandwidth, $|F_\nu / \Delta \nu|^2$. Under this normalization, the amplitude of the individual peaks (in the geometric regime) are exactly given by $|\Delta_j|^{-1}$. Thus, under this normalization, the height of the peaks in delay space give the amplitude of the geometric images, and the location of the peaks give the time delay associated with each image. Note that the top-right panel of Fig.~\ref{fig:ft} only shows the peaks associated with two of the four images. In geometric optics, the imaginary part of $x_j$ for the complex image causes the amplitude to be exponentially suppressed, and so its contribution to the total power is negligible compared to the other images. The other real image is the ``unperturbed" image centred at $\tau = 0$, with $|\Delta_j|^{-1} = 1$. The top-right panel of Fig.~\ref{fig:ft} zooms in on the lensed images so that they are easily resolvable by eye. Both of these images have amplitudes of $|\Delta_j|^{-1} \sim \kappa^{-1} \sim 10^{-3}$. Thus, while the intensity fluctuations shown in the top-left panel may be much larger, $\sim 0.2$, when the oscillations are averaged over (as might occur for an observation with limited frequency resolution), the observed fluctuations could potentially be as small as $10^{-3}$.  

As we describe in Appendix~\ref{sec:appendixB}, one of the main conceptual advantages of using Picard-Lefschetz theory to evaluate the Kirchhoff-Fresnel integral is that even for low-frequencies, one can separate the integral into contributions from a discrete set of images that correspond to the geometric limit as one increases the frequency. The height of the black X's in the right column of Fig.~\ref{fig:ft} correspond to the amplitude of each of the relevant images computed using Picard-Lefschetz theory, shown at the corresponding time delay. In particular, one can see that while the imaginary image is exponentially suppressed in geometric optics, and thus not shown on the top-right panel, as the frequency decreases, the imaginary image becomes significant, and indeed has an amplitude comparable to the real images in the bottom panel. 

Now, while separating the wave-field into distinct contributions from the relevant images is a mathematically well-defined procedure under Picard-Lefschetz theory, we can see that the form of the conjugate wave-field in delay space is qualitatively very different between the geometric and diffractive/perturbative regimes. Specifically, the conjugate wave-field for the perturbative regime is no longer a simple sum of delta functions. While, regardless of frequency, there is always a peak centred at $\tau = 0$ and with height of unity corresponding to the unperturbed image (which always must be present even in the absence of a lens), the lensed images cease to form well-defined peaks in delay space, and instead merge into a single peak with some width. The width of this merged peak increases as the frequency decreases. 

Another way to think of this is by considering two fundamental time-scales in delay space. First, there is the resolution scale given by $(\Delta \nu)^{-1}$. That is, the maximum resolution in the variable $\tau$ for which one can compute the conjugate wave-field is given by the inverse of the bandwidth over which one observes the wave-field itself. The second time-scale one may consider is the relative delay between the lensed images. Consider the two real images (aside from the un-perturbed image), which we will call $x_1$ and $x_2$. Then this relative delay is $\Delta \tau \equiv |T(x_1) - T(x_2)|$. This delay between the lensed images can be thought of as the effective geometric width of the lens in delay space.

The conjugate variable $\tau$, scaled by both of these time-scales, is shown in Fig.~\ref{fig:ft}, with the values of $\tau \Delta \nu$ being shown on the bottom horizontal axis, and the values of $\tau/\Delta \tau$ shown on the top horizontal axis. Note that geometric optics corresponds to the regime when $\Delta \tau \gg (\Delta \nu)^{-1}$. Conversely, the perturbative regime occurs when $\Delta \tau \ll (\Delta \nu)^{-1}$. This can be easily made sense of when we realize that for large $\kappa$, we have $\Delta \tau \sim \kappa$. Thus, $\Delta \tau \sim (\Delta \nu)^{-1}$ when $\epsilon \sim \kappa \Delta\nu \sim 1$. In other words, geometric optics is obtained in the limit where the fundamental resolution limit in delay space is much smaller than the relative time delays of the lensed images. Diffractive/perturbative optics occurs when the relative time delays become un-resolvable in the conjugate wave-field. Observationally, this leads to a rather straightforward way of diagnosing whether or not an observation occurs in the geometric or diffractive regime, since it is often possible to measure the relative time delays of scattered images.

\section{Implications for turbulent plasma lensing}
\label{sec:turbulence}

In this section we will discuss the implications of our results for the study of plasma lensing. Following observations of extreme scattering events (ESEs) in pulsars \citep{Fiedler1987} and the discovery of pulsar scintillation \citep{1968Natur.218..920S,1977ARA&A..15..479R}, both of which are thought to be due to scattering in the interstellar medium (ISM), there has been significant effort to understand wave effects in plasma lensing. The expectation that fast radio bursts (FRBs) will be lensed in a similar way to pulsars has increased the need for a robust understanding of wave optics in lensing. 

Due to the difficulties in computing the Kirchhoff-Fresnel integral, especially for a random ensemble of lenses, studies of scattering phenomena in radio astronomy have generally restricted their attention to different regimes of approximation. In the early days of the study of pulsar scintillation, a diffractive framework for interpreting scintillation observations was developed, i.e. diffractive interstellar scintillation (DISS) \citep{1975ApJ...196..695L, 1977ARA&A..15..479R}. Predictions of so-called refractive interstellar scintillation (RISS) were later confirmed by observations of periodic structures in the secondary spectra of scintillation patterns \citep{1986ApJ...307L..27C, 1997MNRAS.287..739R}. Further confirmation of this refractive ``multi-imaging" effect came in the form of extreme scattering events (ESEs) which exhibited bright caustic structures and were first observed in pulsars in 1993 \citep{1993Natur.366..320C}. Although, as we have stressed, the phenomena of DISS and RISS are, in principle, purely observationally defined, it has often been assumed that RISS is described by multiple imaging in the geometric optics regime, whereas DISS is a purely wave phenomenon \citep{1986ApJ...310..737C}. This assumption is important, as it leads to the widely utilized relationship that relates the observed scattering angle to the size of the plasma inhomogeneities, $\alpha^* \sim \lambda / a$. This relation has been used to estimate sizes of turbulent eddies in the ISM \citep[see e.g.][]{2005MNRAS.358..270W, 2011A&A...534A..66L, 2016MNRAS.462.3115D}. We argue that our results here, made possible by a novel framework for exactly computing the Kirchhoff-Fresnel integral, suggest that the applicability of a diffractive framework to scintillation observations has potentially been overestimated. Indeed, in recent decades, following the discovery of parabolic arcs and inverted arclets in many pulsar secondary spectra \citep{2001ApJ...549L..97S, 2010ApJ...708..232B}, a purely refractive framework has been successfully applied to explain a surprising variety of scintillation phenomena than one might have expected given the early emphasis on diffractive phenomena \citep{2004MNRAS.354...43W, 2006ApJ...640L.159G, 2021MNRAS.500.1114S}. Indeed, there are some suggestions that all scattering phenomena of relevance in radio astronomy are well within the refractive regime \citep{2014MNRAS.442.3338P}.

In order to compare our results with the turbulent plasma lensing literature, we will first re-write the Kirchhoff-Fresnel integral using the dimensionful parameters that appear in plasma lensing. Eq.~\eqref{eq:Kirchhoff-Fresnel} becomes
\begin{equation}
    F(y) \propto \int \exp \Big\{ i\frac{\omega}{\overline{d}c}\Big[ \frac{(\hat{x} - \hat{y})^2}{2} + \frac{\overline{d} c e^2 \Sigma_e(\hat{x})}{2 m_e \epsilon_0 \omega^2} \Big] \Big\} d\hat{x},
    \label{eq:Kirchhoff-Fresnel-dimensionful}
\end{equation}
where $\omega$ is the angular frequency of the source radiation, $\overline{d} = d_{sl} d_l / d_s$ is the reduced distance, $\hat{y} = (\hat{x}_s d_l + \hat{x}_o d_{sl} ) / d_s$ is a weighted average of the transverse displacement of the source and observer, and $\Sigma_e$ is the projected electron column density of the plasma in the lens plane. Fig.~\ref{fig:lensing_geometry} shows the distances and the coordinates involved in this parametrization. Note that this form of the phase induced by a plasma is valid only when the angular frequency of the light is large compared to the plasma frequency, $\omega_p^2 = n_e e^2 / m_e \epsilon_0$. However, for typical electron densities in the ISM, $n_e \sim 1\,{\rm cm}^{-3}$, the plasma frequency is $\omega_p \sim 1\,{\rm kHz}$. For radio sources such as FRBs and pulsars, the plasma frequency is always many orders-of-magnitude smaller than any frequency of interest.

Now, in order to compare this with our dimensionless analysis, we re-scale the coordinates by $a$, which we take to be the size of the lens. That is, we take $\hat{x} \to x = \hat{x}/a$,  $\hat{y} \to y = \hat{y}/a$, and $\Sigma_e (\hat{x}) \equiv \Sigma_0 \psi(x)$, where $a$ is defined so that $\psi''(0) \approx -2$. Defined in this way, $\Sigma_0$ is the amplitude of the plasma surface density variation. Thus, Eq.~\eqref{eq:Kirchhoff-Fresnel-dimensionful} reduces to Eq.~\eqref{eq:Kirchhoff-Fresnel}, with
\begin{align}
    \nu &= \frac{\omega a^2}{c \overline{d}} = \frac{a^2}{R_F^2},
    \label{eq:nu_dimensions}
    \\
    \kappa &= \frac{\overline{d} e^2 \Sigma_0}{2 a^2 m_e \epsilon_0 \omega^2},
    \label{eq:alpha_dimensions}
\end{align}
where $R_F \equiv (\lambda \overline{d} / 2\pi)^{1/2}$ is the dimensionful Fresnel scale. Thus, as we stated before, the condition that $\nu$ be large is equivalent to the Fresnel scale being small relative to the size of the lens. For plasma lensing, we obtain
\begin{equation}
    \epsilon = \frac{e^2 \Sigma_0}{2 m_e c \epsilon_0 \omega}.
    \label{eq:plasma_epsilon}
\end{equation}
Note that since the convergence scales with the inverse frequency squared in plasma lensing, we have $\epsilon \sim \omega^{-1}$. Thus, geometric optics ($\epsilon \gg 1$) is obtained in the low frequency limit, and diffractive optics ($\epsilon \ll 1$) in the large frequency limit. 

The primary difficulty in comparing our results to the previous literature is that much of the study of wave optics in plasma lensing was performed for turbulent lenses. That is, whereas we consider a smooth $\psi(x)$ with a single peak (and later an ensemble of such lenses in Section~\ref{sec:strong}), studies of turbulent plasma lensing consider a Gaussian random field as the lensing potential. The primary goal of these studies is not to compute the observed light curve for a particular lens, but to compute the statistical properties of the electric field at the observer given some power spectrum defining the lens. For example, \citet{Rickett1990} reviews the study of turbulent plasma lenses with power spectra of the form
\begin{equation}
    P(k) \propto (k^2 + k^2_\mathrm{outer})^{-\beta/2} \exp (-k^2 / k^2_\mathrm{inner}).
    \label{eq:power_spectrum}
\end{equation}
Between the inner and outer scales set by $k_\mathrm{inner}$ and $k_\mathrm{outer}$, this power spectrum is a simple power law, $P(k) \propto k^{-\beta}$, and, in particular, is Kolmogorov for $\beta = 11/3$. Outside of these scales, the power spectrum falls rapidly to zero. 

The phenomena of DISS and RISS are observed to occur on different spatial scales, separated by the Fresnel scale. That is, the intensity fluctuations are described as DISS below the Fresnel scale, whereas intensity fluctuations above the Fresnel scale are said to be due to RISS. In order to explain the emergence of diffractive and refractive interstellar scintillation, \citet{1986ApJ...310..737C} separate the lens potential into two distinct parts: a refractive part, $\psi_r$, and a diffractive part, $\psi_d$, such that $\psi = \psi_r + \psi_d$. The total observed flux is then given by
\begin{equation}
    F \propto \sum_{x_r} \mu_r \int dx e^{\nu \big[ \frac{1}{2} (x-x_r)^2 + \psi_d(x) \big]},
    \label{eq:Cordes}
\end{equation}
where the refractive images, $x_r$, are given by the solutions to the lens equation for the refractive part of the phase, and $\mu_r$ is the geometric optics magnification of that image. That is, in this framework, the refractive part of the phase gives rise to multiple images described by geometric optics, and the effect of the diffractive part of the phase is to modulate the intensity of these images. It is often presumed that the evaluation of the integral in Eq.~\ref{eq:Cordes} cannot be further simplified using geometric optics and so the diffractive part represents a purely wave optical effect. Strong diffractive scintillation is said to occur when the phase variation due to the diffractive part is large, i.e. when $\langle |\nu \psi_d|^2 \rangle \gtrsim 1$.  

A quantity of central importance that was introduced to the study of turbulent plasma scattering is the coherence scale, $s_0$, which is defined as the average lateral separation in the lens plane across which the total phase, $\nu \psi$, undergoes a change of one radian \citep{Rickett1990}. The strength of scintillation is then characterized by the quantity
\begin{equation}
    u = \frac{R_F}{s_0},
    \label{eq:u}
\end{equation}
such that when $u < 1$ the amplitude of the intensity fluctuations predicted by the perturbative expansion are weak and when $u > 1$ they are strong. Essentially, when the phase variations are large, $\langle |\nu \psi|^2 \rangle \gtrsim 1$, the coherence scale is small relative to the Fresnel scale. The parameter $u$ encodes the same information as the parameter $\epsilon$ in our rational lens model: when $u$ (or $\epsilon$) is much greater than unity, the phase variation is large and the lensing can be said to be ``strong". Much of the early work on turbulent plasma lensing was concerned with identifying the observational phenomena of DISS and RISS with different regimes within this mathematical framework. Using heuristic arguments, it was argued that the Fresnel scale sets the relevant scale between the diffractive and refractive regimes. These arguments can be found in more detail in \citet{Rickett1990} and \citet{handbook}, but we will summarize them here for the sake of completeness.

\begin{figure}
    \centering
    \includegraphics[width=\columnwidth]{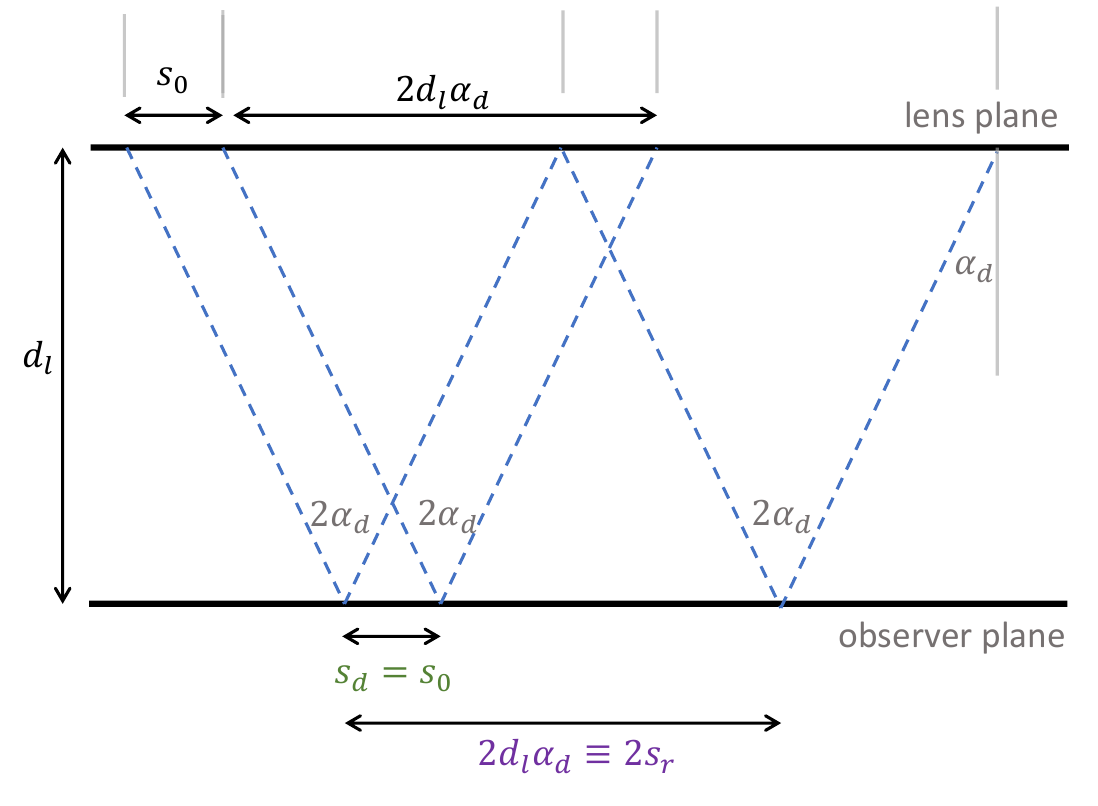}
    \caption{Diagram of the turbulent plasma scattering picture for a source at infinity. Diffractive intensity variations at an observer come from waves emanating from within a angle of $2 \alpha_d$, where $\alpha_d$ is the average scattering angle. The diffractive scale, $s_d$, is taken to be roughly equal to the coherence scale, $s_0$, which is the average lateral distance over which the phase along a straight line path from source to observer varies by one radian. The refractive scale, $s_r$, is taken to be $\overline{d} \alpha_d$, as observers separated by larger than this have no overlap between their scattering disks.}
    \label{fig:scattering_diagram}
\end{figure}

From the definition of the coherence scale, $s_0$, one can define the diffractive scattering angle, $\alpha_d \sim \lambda/s_0$, which is roughly the average bending that light rays undergo as they pass through the plasma \citep{1967ApJ...147..433S, Rickett1990}. It is easiest to see how this quantity relates to the diffractive and refractive distance scales defined in the literature for the case where the source is infinitely far away, $d_s \to \infty$, a diagram of which is shown in Fig.~\ref{fig:scattering_diagram}. If we consider the electric field at the observer to be a sum of point sources located along the lens (the Huygens-Fresnel principle), then the observed diffractive intensity variations come only from the waves within the angular spectrum of width $2 \alpha_d$, which is the total angular size of the scattering disk. When the observer is laterally displaced by $s_0$, the relative phase of the incoming waves are substantially shifted from the original observer position (by roughly one radian), resulting in significant differences in the observed intensity across this scale. Thus, the diffractive scale, $s_d$, is taken to be roughly equal to the coherence scale, $s_d \approx s_0$. Now, intensity variations due to refraction (i.e. due to the bending of multiple light rays into the line of sight of the observer) must occur on distance scales larger than $2 d_l \alpha_d$. This is because two observers separated by a distance less than $d_l \alpha_d$ will have overlapping scattering disks, as shown in Fig.~\ref{fig:scattering_diagram}. Thus, we can define the refractive scale $s_r \equiv d_l \alpha_d$, such that any significant intensity variations on scales larger than this must be due to multiple images being refracted into the line of the sight of the observer. It follows from straightforward geometry that in the case where $d_s$ is finite, then the angular size of the scattering disc becomes $2 \alpha_d d_{sl} / d_s$ and the diffractive and refractive scales are $s_d = s_0$ and $s_r = \overline{d} \alpha_d$.

\begin{figure}
    \centering
    \includegraphics[width=\columnwidth]{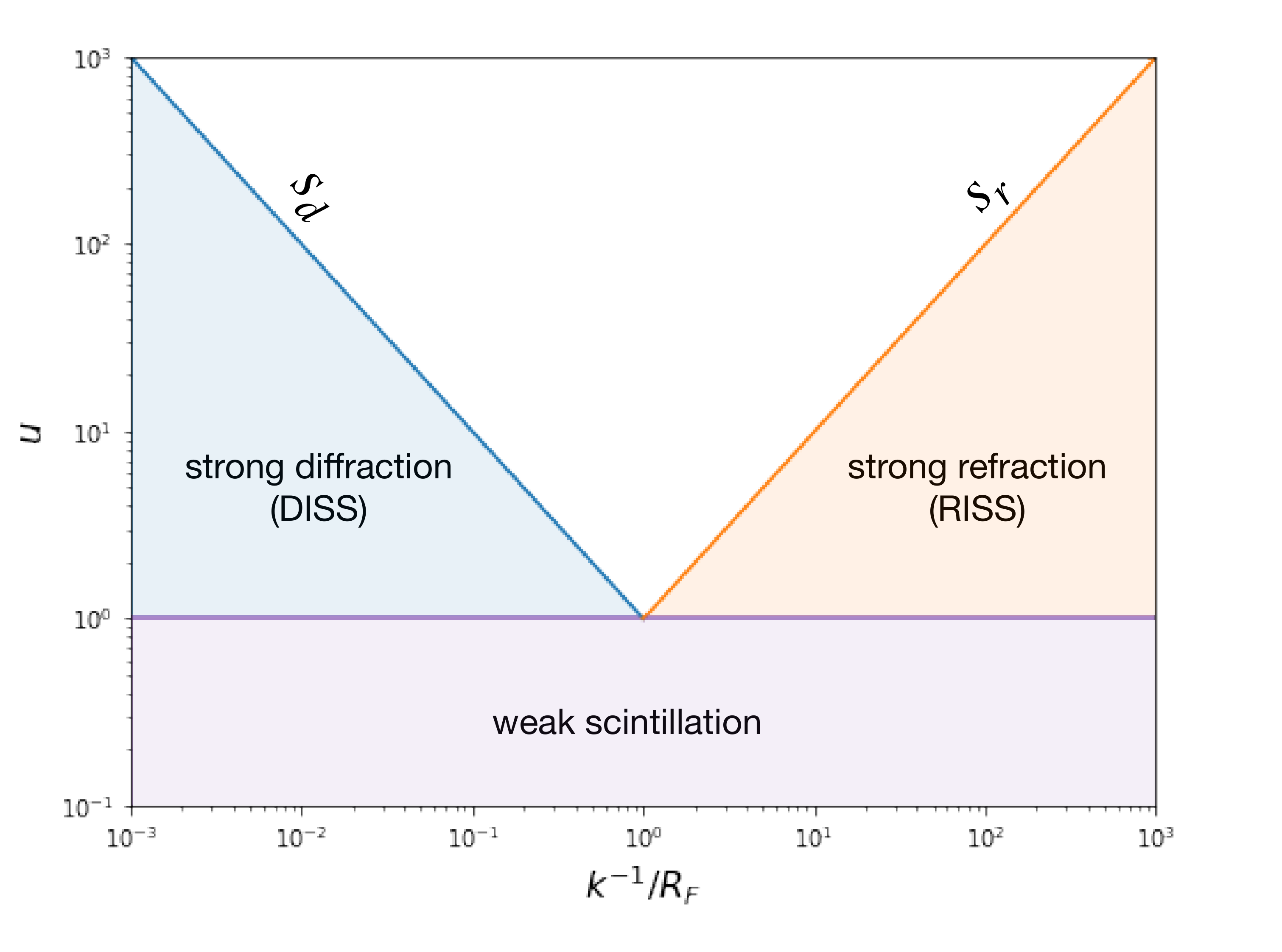}
    \caption{Summary of the regimes described in early turbulent plasma lensing literature. The statistics of the observed electric field is taken to be described by diffractive interstellar scattering (DISS) for spatial scales less than $s_d = R_F / u$ and by refractive interstellar scattering (RISS) for spatial scales greater than $s_r = u R_F $. The quantity $u$ plotted on the $y$-axis sets the predicted strength of the diffractive intensity variations, where they are much less than unity for $u < 1$ and greater than unity for $u > 1$. The $x$-axis is the spatial scale of intensity variations $k^{-1}$ normalized by the Fresnel scale. Note that this description is in contrast with our lens model, for which the refractive regime can be made to hold at arbitrarily small scales when the lens strength is large.}
    \label{fig:lit_diag}
\end{figure}

Altogether, in terms of the Fresnel scale and $u$, one can write the diffractive and refractive scales as \citep{Rickett1990, handbook}
\begin{align}
    s_r &= u R_F, \\
    s_d &= \frac{R_F}{u}.
\end{align}
Fig.~\ref{fig:lit_diag} summarizes this picture of turbulent plasma lensing that arises from this description. When $u < 1$, the intensity fluctuations are weak. When $u > 1$, the intensity fluctuations are strong, and are split into distinct diffractive and refractive regimes which are identified with DISS and RISS. The Fresnel scale sets the boundary between diffractive and refractive lensing, but as $u$ gets large, the region around the Fresnel scale for which neither approximations hold gets larger. Fig. 1 of \citet{Rickett1990}, which shows a schematic of the power spectrum of the observed intensity fluctuations as a function of scale, also presents this picture of lensing where the Fresnel scale sets the boundary between two distinct regimes. 

Our results, however, are at odds with this picture. In the framework we have just described, DISS refers to a regime in which the intensity variations are large ($u > 1$), and the observed flux comes from an angular region on the sky given by $2 \alpha_d d_{sl} / d_s$ where the scattering angle is $\alpha_d \sim \lambda / s_0$. In contrast, in Section~\ref{sec:strong} we argued that for an ensemble of rational lenses, whenever the scattering angle was given by $\alpha^* \sim \lambda / a$, where $a$ is the physical size of the individual lenses, the scintillation was necessarily weak. In other words, we find that diffractive scintillation is automatically weak scintillation (i.e. it is identified with the perturbative regime of optics), and strong scintillation can only arise deep in the refractive regime. Moreover, we find that, for the isolated rational lens, rather than the Fresnel scale setting the transition point between diffractive (perturbative) and refractive (geometric) optics, the dimensionless parameter $\epsilon = \kappa \nu$ does. Indeed, as can be seen in Fig.~\ref{fig:transition}, refractive optics can be made to hold at arbitrarily small scales well below the Fresnel scale.

It is, of course, challenging to make inferences regarding turbulent plasma lensing from our results for the rational lens. A turbulent lens is an ensembles of random phases over many scales with a given power spectrum, whereas the rational lens we consider is a single, isolated entity. It is particularly challenging to relate the quantity $s_0$ to the parameters of our isolated lens model, since this is a statistically defined quantity. However, there is a regime in which the evenly spaced ensemble of rational lenses that we describe in Section~\ref{sec:ensemble} can be compared to a turbulent plasma with power spectrum given by Eq.~\ref{eq:power_spectrum}. Namely, our ensemble model is qualitatively similar to a turbulent model where the lensing is dominated by structures close to the inner scale of the turbulence (i.e. where $s_0 \sim k_{\rm inner}^{-1}$). Now, in order to relate the physical inner scale with the relevant distance scale in our model, $a$, we note that we are primarily concerned with understanding the implications of our results for the phenomenon of DISS, which is defined to be a strong phenomenon ($u \gtrsim 1$ in the language of turbulent plasma lensing). In the language used in this paper, ``strong" is identified with $\epsilon \gtrsim 1$. When $\epsilon = 1$, then the width in the lens plane over which the induced phase $\epsilon \psi(x)$ changes by one radian is precisely $a$. Thus, in this transitory regime, $\epsilon \sim 1$, where we can most easily compare our results to a turbulent plasma lensing framework, the physical lens scale $a$ is the same as the coherence scale $s_0$. In other words, the scattering angle for the regime we identify with diffractive optics is given by $\alpha^* \sim \lambda / s_0$, which is what is assumed to hold for DISS. We take this as additional justification for the identification of diffractive optics with this regime. Our results, therefore, suggest that an observation of strong scintillation (modulation index of order unity) is mutually incompatible with the assumption of DISS that the scattering angle is given by $\alpha^* \sim \lambda / s_0$. At the very least, our results suggest that this relation is crucially dependent on the power spectrum of the turbulence, and may not be true in general. Specifically, we have argued that it fails for lensing dominated by structures close to the inner scale. 

\subsection{Scattering in the Black Widow pulsar, B1957+20}
\label{sec:B1957}

Our arguments in this section have been purely heuristic. In order to make them more concrete, it will be necessary to apply our analysis to the more general case of a Gaussian random field lens: work which will appear in a future publication. However, we stress that this is not merely a semantic argument. Whether or not the relation $\alpha^* \sim \lambda / s_0$ holds in any regime of strong scintillation has practical implications for the physical inferences made from observations. As a concrete example, consider the plasma lensing of the Black Widow pulsar, B1957+20, presented in \citet{2022arXiv220813868L}. Pulsar B1957+20 is a binary system where the pulsar is lensed by the plasma outflow from its companion \citep{2018Natur.557..522M}. While the lensing is not, strictly speaking, interstellar scintillation, which is due to plasma inhomogeneities in the ISM, it is nevertheless an instructive example to consider, as many of the relevant lensing parameters are known a priori. For example, since the lensing is known to be associated with the companion's outflow, and the orbital parameters of the system are known, the relative velocity between the source and lens and the distance to the lens screen can be inferred. As such, we can directly test whether or not the DISS framework gives the correct answer for the size of the lensing structures. As shown in \citet{2022arXiv220813868L}, the lensing of PSR B1957+20 is split into two distinct regimes: an apparent weak lensing regime and a strong lensing regime. The apparent weak lensing regime is referred to in this way because it is well-described by a simple linear model; i.e. the magnification is fully determined by spatial derivatives of the plasma density around the unperturbed line of sight (see \citet{2022arXiv220813868L} for a more complete discussion). In other words, according to the traditional turbulent plasma lensing framework, this regime would be considered diffractive lensing, and the size of the plasma inhomogeneities should be given by $s_0 \sim \lambda / \alpha^*$. The wavelength of the observation is $\lambda = 1\,{\rm m}$ and the scattering angle can be inferred from the observed scattering timescale, $\tau_{\rm scattering} \sim \alpha^2 R / c \sim 0.1\,{\rm ms}$, where $R \approx 3\,R_{\odot}$ is the orbital separation between the pulsar and its companion. It follows from this that the predicted size of the inhomogeneities in the DISS framework would be $s_0 \sim 300\,{\rm m}$. However, we can directly measure the size of the inhomogeneities since we know both the timescale of the lensing events (see the autocorrelation timescale in Fig.~2 of \citet{2022arXiv220813868L}) and the relative velocity between the source and lens. Combining these two pieces of information gives a measurement of the size of the lensing structures of $v t \sim 300\,{\rm km} \, {\rm s}^{-1} \cdot 1\,{\rm s} \sim 300\,{\rm km}$, which is a full three orders of magnitude larger than the size predicted by the DISS framework. Thus, we suggest that even though the lensing may have characteristics that have been previously associated with ``diffractive" lensing, one must be careful when applying the diffractive formula $\alpha^* \sim \lambda / s_0$.

\section{Typical values of dimensionless parameters}
\label{sec:values}

In this section, we compute the dimensionless parameters $\nu$, $\kappa$, and $\epsilon$ for typical physical parameters that may occur in plasma lensing and gravitational lensing.

\subsection{Plasma lensing}

The study of wave optics in astrophysical plasma lensing is particularly relevant to pulsar observations, as pulsars are effectively coherent point sources. Pulsars are radio sources with frequencies of order $~{\rm GHz}$, and are found within the galaxy. Thus, we can expect typical distances to be $\overline{d}\sim1\,{\rm kpc}$. Pulsars have been observed to undergo extreme scattering events (ESEs) due to lensing by roughly AU-scale features in the ISM, with excess electron column densities of order $\sim0.01\,{\rm pc\,cm}^{-3}$ \citep{2015ApJ...808..113C,2018MNRAS.474.4637K}. Thus, using Eqns.~\ref{eq:nu_dimensions} and ~\ref{eq:alpha_dimensions}, we obtain
\begin{align}
    \nu &= 1.5 \times 10^4\,\Big(\frac{f}{1\,\mathrm{GHz}}\Big) \Big(\frac{a}{1\,\mathrm{AU}} \Big)^2 \Big( \frac{\overline{d}}{1\,\mathrm{kpc}} \Big)^{-1}, \\
    \kappa &= 17\,\Big(\frac{\Sigma_0}{0.01\,\mathrm{pc\,cm}^{-3}}\Big)\Big( \frac{\overline{d}}{1\,\mathrm{kpc}}\Big) \Big(\frac{a}{1\,\mathrm{AU}} \Big)^{-2} \Big(\frac{f}{1\,\mathrm{GHz}}\Big)^{-2}, \\
    \epsilon &= 2.6 \times 10^5\,\Big(\frac{\Sigma_0}{0.01\,\mathrm{pc\,cm}^{-3}}\Big) \Big(\frac{f}{1\,\mathrm{GHz}}\Big)^{-1}.
    \label{eq:epsilon_plasma}
\end{align}
Consistent with ESE observations, $\kappa \gg 1$, resulting in multiple real images and large intensity fluctuations. Moreover, we note that $\epsilon \gg 1$, suggesting that ESEs occur firmly in the geometric regime. 

Note that the distance parameters $\overline{d}$ and $a$ both cancel out in the expression for $\epsilon$. Thus, the factor which determines whether or not lensing occurs in the geometric or diffractive/perturbative regime depends only on the excess electron surface density and the frequency. In particular, we can use Eq.~\eqref{eq:epsilon_plasma} to argue that pulsar scintillation generically occurs in the geometric regime, as $\epsilon \sim 1$ only occurs when the excess electron surface density is $\Sigma_0 \sim 10^{-8}\,{\rm pc\,cm}^{-3}$. 

\subsection{Gravitational lensing}

In gravitational lensing, the surface mass density, as opposed to the electron column density, sets the amplitude of the lens potential. Moreover, the amplitude only depends on the surface mass density, and does not depend on frequency. In gravitational lensing, one is also often concerned with cosmological distances, and so the distances in $\overline{d} = d_{sl}d_l / d_s$ must be replaced by angular diameter distances, and the frequency $\omega$ must include a redshift correction, $\omega \to (1+z)\omega$, where $z$ is the redshift of the lens \citep{Schneider}. In other words, since the lensing time-delays depend on the geometric difference between paths through the lens, the relevant cosmological distance measure is the angular diameter distance. Moreover, the relevant frequency that determines the phase offset induced by the lens is the frequency at the lens, $\omega (1 + z)$. Otherwise, the formalism is exactly the same as plasma lensing.

We may wish to know the values of the dimensionless parameters in the case of lensing by a galaxy at cosmological distances. Thus, the lens has a size of order $\sim\,10\,{\rm kpc}$ and has a distance of order $\sim1\,{\rm Gpc}$. Since fast radio bursts (FRBs) are the only coherent sources at cosmological distances, we will also be primarily concerned with radio frequencies. Given these typical parameters, the dimensionless frequency $\nu$ can be computed:
\begin{equation}
    \nu \sim 6.5\times 10^{16}\,(1+z)\Big(\frac{f}{1\,\mathrm{GHz}}\Big) \Big(\frac{a}{10\,\mathrm{kpc}} \Big)^2 \Big( \frac{\overline{d}}{1\,\mathrm{Gpc}} \Big)^{-1}.
    \label{eq:nu_grav}
\end{equation}
Here, $f$ is the observing frequency. The relevant frequency for computing $\nu$ is the frequency at the lens, $f (1 + z)$; hence the inclusion of the redshift term.

Many lens profiles have been used to model lensing by galaxies. One simple model is the Plummer model for which it is straightforward to re-parametrize the convergence in terms of the total lens mass. One obtains that the convergence is given by $\kappa = 2 G M \overline{d} / a^2 c^2$ \citep{2006MNRAS.368.1362W}. 
\begin{equation}
    \kappa \sim 1\,(1+z)^{-1}\Big(\frac{M}{10^{12}\,M_\odot}\Big)\Big( \frac{\overline{d}}{1\,\mathrm{Gpc}}\Big) \Big(\frac{a}{10\,\mathrm{kpc}} \Big)^{-2} .
    \label{eq:kappa_grav}
\end{equation}
Combining Eqs.~\ref{eq:nu_grav} and ~\ref{eq:kappa_grav}, we obtain
\begin{equation}
    \epsilon = 6.5\times10^{16}\Big(\frac{M}{10^{12}\,M_\odot}\Big) \Big(\frac{f}{1\,\mathrm{GHz}}\Big),
\end{equation}
which is generically much larger than unity.

\section{Conclusion}
\label{sec:conclusion}

Through the introduction of the mathematical framework of Picard-Lefschetz theory to the field of astrophysical lensing, it has now become possible to evaluate the Kirchhoff-Fresnel diffraction integral for a wide variety of situations. Previously, theoretical treatments of the study of wave optics in astrophysical contexts relied on various approximation schemes. Exact methods of computing diffraction integrals allow us to quantitatively evaluate longstanding assumptions in the field.

In this work, we have studied a simple lens model to assess when the perturbative approximation holds, and when the geometric limit of optics holds. Through our study of the simple rational lens, we identify the perturbative regime with ``diffractive" optics and the geometric regime with ``refractive" optics. We find that, for the lens model we consider, there is no overlap in the regime of applicability of these two methods, and that the region of parameter space where neither method is valid is limited. That is, optics is generically well described by either diffractive or refractive optics. We find that the spatial scale that separates these two regimes is given by the Fresnel scale divided by the square-root of the convergence, $R_F/\sqrt{\kappa}$. That is, for a lens with convergence $\kappa$ and size $a$, refractive optics holds when $a$ is larger than $R_F/\sqrt{\kappa}$, and diffractive optics holds when it is smaller. Previously, it has been assumed that the Fresnel scale alone sets the scale at which diffractive optics is valid. We have also argued that large intensity variations do not generically occur in the diffractive regime. This has important implications for the study of pulsar scintillation. 

In this work, we have focused on a simple, one-dimensional lens model. The highly non-linear nature of the Kirchhoff-Fresnel integral means that it is not straightforward to apply these results to more complicated lensing systems. Further detailed study of systems of interest, for example, turbulent plasma lensing, is needed. In particular, a similar analysis of the Gaussian random lens is a natural next step.

\section*{Data Availability}
No new data were generated or analysed in support of this research.

\section*{Acknowledgements}
We thank Fang Xi Lin, Dan Stinebring, and Barney Rickett for useful  discussions. We receive support from Ontario Research Fund—research Excellence Program (ORF-RE), Natural Sciences and Engineering Research Council of Canada (NSERC) [funding reference number RGPIN-2019-067, CRD 523638-18, 555585-20], Canadian Institute for Advanced Research (CIFAR), Canadian Foundation for Innovation (CFI), the National Science Foundation of China (Grants No. 11929301),  Thoth Technology Inc, Alexander von Humboldt Foundation, and the Ministry of Science and Technology(MOST) of Taiwan(110-2112-M-001-071-MY3). Computations were performed on the SOSCIP Consortium’s [Blue Gene/Q, Cloud Data Analytics, Agile and/or Large Memory System] computing platform(s). SOSCIP is funded by the Federal Economic Development Agency of Southern Ontario, the Province of Ontario, IBM Canada Ltd., Ontario Centres of Excellence, Mitacs and 15 Ontario academic member institutions.
 Cette recherche a \'{e}t\'{e} financ\'{e}e par le Conseil de recherches
en sciences naturelles et en g\'{e}nie du Canada (CRSNG), [num\'{e}ro de
r\'{e}f\'{e}rence 523638-18,555585-20 RGPIN-2019-067]. JF is supported in part by the Higgs Fellowship.

\bibliographystyle{mnras_sjf}
\bibliography{biblio} 



\appendix

\section{Picard-Lefschetz analysis of the rational lens}
\label{sec:appendixB}

The purpose of this work has been to evaluate the regimes of validity of the refractive and diffractive regimes of optics through explicit comparison with the exact result of the Kirchhoff-Fresnel integral. In particular, we studied the simple rational lens, $\psi = 1/(1+x^2)$, as an example. Our ability to exactly compute the Kirchhoff-Fresnel integral for this lens is enabled by the tools of the mathematical framework of Picard-Lefschetz theory, which was re-introduced to the study of oscillatory integrals in physics by \citet{2010arXiv1001.2933W}, and has subsequently been applied to the study of lensing \citep{job_pl, Jow2021, 2021MNRAS.506.6039S}. \citet{2010arXiv1001.2933W} and \citet{job_pl} give a detailed description of the theoretical and numerical application of Picard-Lefschetz theory to oscillatory integrals, but we will give a brief summary here which follows the description of the theory in \citet{Jow2021}.

The goal is to compute integrals of the form
\begin{equation}
    I = \int_{\mathbb{R}} dx e^{i S(x; \mathbf{\xi})},
    \label{eq:gen_osc_int}
\end{equation}
where $\mathbf{\xi}$ is a set of parameters that fix the phase function $S(x;\mathbf{\xi})$. Picard-Lefschetz theory is an application of Cauchy's theorem for the deformation of integration domains and guarantees the existence of a surface in the complex plane, $\mathbb{C}$, that is a continuous transformation of the original integration domain, $\mathbb{R}$, such that the integrand is localized and non-oscillatory.

To see how this works, we define the function $h(z) = \text{Re}\{i S(z)\}$. We assume that $S(z)$ is analytic, and we have replaced the integration variable $x$ with the variable $z$ to denote the analytic continuation of the integrand to the complex plane. It follows from the Cauchy-Riemann equations that the critical points of the function $h(z)$ are also the critical points of $S(z)$. If the critical points of $h(z)$ are non-degenerate, then $h(z)$ is a Morse function, and it follows from a basic result of Morse theory that every critical point has a Morse index of $n$, where $n$ is the dimension of the complex manifold. Since we are considering the one dimensional case, this means that every critical point of $h(z)$ has a Morse index of unity, or, in other words, is a saddle point. Thus, every critical point $z_j$ of $h(z)$ has associated with it a contour of steepest descent, $\mathscr{J}_j$, with a dimension of one, and a contour of steepest ascent, $\mathscr{U}_j$, of equivalent dimension. 

The contours of steepest descent are determined by the flow equations
\begin{equation}\label{eq:gradflow}
    \frac{dz}{dt} = \mp \frac{d (i\bar{S})}{d\bar{z}},\quad     \frac{d\bar{z}}{dt} = \pm \frac{d (iS)}{d z},
\end{equation}
where $t$ parameterizes the contour, $z(t)$. When the sign in front of the right-hand side of the first of Eq.~\eqref{eq:gradflow} is negative (and the sign of the second is positive), the equations determine the trajectories of steepest descent. When the signs are reversed, the equation determines the trajectories of steepest ascent. The surface $\mathscr{J}_j$ associated with $z_j$ is a manifold determined by the set of initial points for which solutions to the steepest ascent equation terminate at $z_j$ as $t \to \infty$. Hence, the surface $\mathscr{J}_j$ is a contour of descent, since the direction pointing away from the critical point $z_j$ along the surface $\mathscr{J}_j$ is a direction of decreasing $h$.  

One can show that along each of the contours of steepest descent, $\mathscr{J}_j$, the quantity $\mathrm{Im}\{ iS(z) \}$ is constant:
\begin{align}
\frac{\mathrm{d} \text{Im}[iS]}{\mathrm{d}t}= \frac{1}{2i} \frac{\mathrm{d}(i S + i \bar{S})}{\mathrm{d}t}
= \frac{1}{2i}\left(\frac{\partial (iS)}{\partial z} \frac{\mathrm{d} z}{\mathrm{d}t}+\frac{\partial (i\bar{S})}{\partial \bar{z}} \frac{\mathrm{d} \bar{z}}{\mathrm{d}t}\right)=0.
\end{align}
Moreover, the total integrand exponentially approaches $0$ away from the critical point. This means that the integrand is localized and non-oscillatory along these surfaces, which we call the ``Lefschetz thimbles". Picard-Lefschetz theory guarantees that the integral along the original domain, $\mathbb{R}$, is equal to a sum of integrals along the Lefschetz thimbles. Explicitly,
\begin{align}
I = \sum_{j} N_j \int_{\mathscr{J}_j} dz e^{i S(z; \mathbf{\xi})},
\end{align}
where $N_j = \langle \mathbb{R}, \mathscr{U}_j \rangle$ is the intersection number, which counts the number of times the steepest ascent contour, $\mathscr{U}_j$, intersects the original integration domain, $\mathbb{R}$, with sign given by the relative orientations of the two curves. Thus, a thimble, $\mathscr{J}_j$, contributes to the integral if and only if the contour of steepest ascent, $\mathscr{U}_j$, for the corresponding critical point, $z_j$, intersects the original integration domain. In order to compute the integral for a given set of parameters, $\xi$, one must find the complex critical points of $z_j$, as well as the surfaces of steepest ascent and descent associated with each critical point, which can be achieved by solving the flow equations. Note that here we have given a brief description of Picard-Lefschetz theory for one-dimensional oscillatory integrals, but the theory may be extended to arbitrary dimension \citep[see][]{2010arXiv1001.2933W,job_pl}

\begin{figure}
    \centering
    \includegraphics[width=\columnwidth]{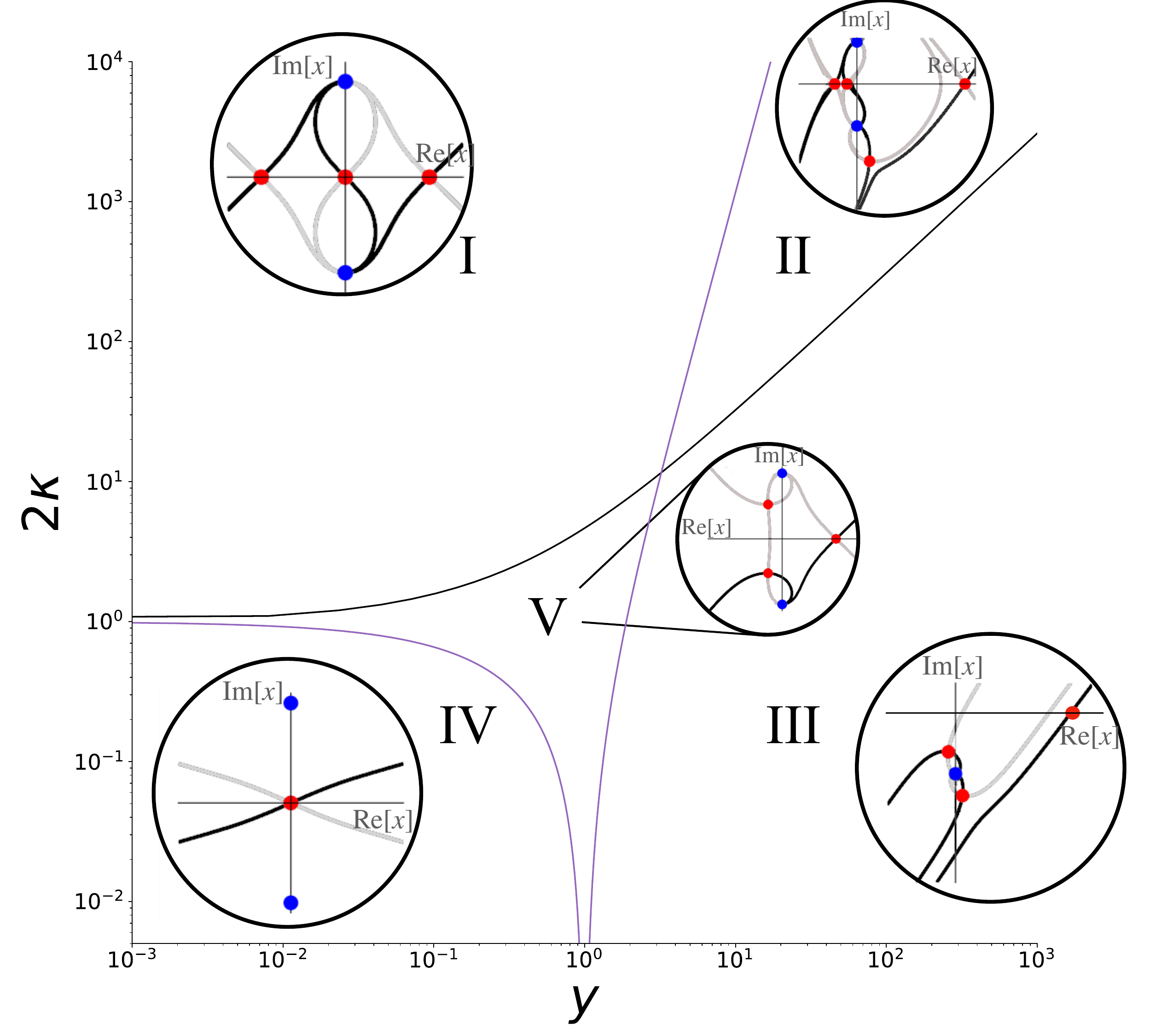}
    \caption{Lefschetz thimble topologies for the rational lens, $\psi = 1/(1+x^2)$. The black and purple lines show the position of the caustics and Stokes lines in lens parameter space, which separate the regions (labeled I, II, III, IV, and V) with distinct thimble topology. The encircled diagrams show the topology of the relevant thimbles in the complex domain for each of these regions. That is, the encircled diagrams show examples of the optimal contours of integration prescribed by Picard-Lefschetz theory after analytically continuing the integral into the complex domain, $x \to x = {\rm Re}[x] + i {\rm Im}[x]$. The red dots correspond to the relevant critical points, and the black and grey curves correspond to the surfaces of steepest descent and ascent associated with each critical point. The blue dots correspond to the poles of the analytically continued phase function.}
    \label{fig:topology}
\end{figure}

The topology of the Lefschetz thimbles in the complex domain varies smoothly with the parameters $\xi$ except at caustics and Stokes lines. Caustics are points in parameter space where a set of real critical points become degenerate. Lensing near caustics has been widely studied in both the geometric and diffractive regimes \citep{Nye}. Stokes lines are points at which the number of relevant complex critical points changes, and occur when the Lefschetz thimbles of multiple critical points overlap (see \citet{2010arXiv1001.2933W} and \citet{Jow2021} for a more detailed description of Stokes lines). Together, the caustics and Stokes lines form separate regions in parameter space where the Lefschetz thimble topology is distinct. Fig.~\ref{fig:topology} shows the different thimble topologies for the rational lens. The black and purple lines are the caustics and Stokes lines, respectively, which separate the parameter space into five distinct regions. The encircled diagrams for each region show the relevant thimbles in the complex domain. The red dots correspond to the relevant critical points, and the black and grey curves correspond to the surfaces of steepest descent and ascent associated with each critical point. The blue dots correspond to the poles of the analytically continued phase function. As defined in Eq.~\eqref{eq:Kirchhoff-Fresnel}, the integral would be performed over the real axis and is conditionally convergent. However, Picard-Lefschetz theory  guarantees that the integral over the relevant Lefschetz thimbles (the black curves in the encircled diagrams) is not only equal to the original integral over the real axis, but is absolutely convergent. 

Note that the thimble topologies for the rational lens only depend on the parameters $\kappa$ and $y$, as $\nu$ is simply an overall factor multiplying the exponent of the integrand in the Kirchhoff-Fresnel integral (Eq.~\eqref{eq:Kirchhoff-Fresnel}). Thus, $\nu$ does not affect the solution to the flow equations, and thereby does not affect the thimble topology. Positive values of $y$ are shown on the horizontal axis of Fig.~\ref{fig:topology} as the result of the Kirchhoff-Fresnel integral is symmetric about $y=0$ for the rational lens. The vertical axis is taken to be $2 \kappa$, as the cusp catastrophe occurs at $2 \kappa = 1$.

One of the most powerful aspects of Picard-Lefschetz theory as applied to lensing is that it allows for the separation of the total integral into contributions associated with a discrete set of critical points, $\{ {\bm z}_i \}$, by performing the integral along the associated Lefschetz thimbles separately. Moreover, these critical points are precisely the images of geometric optics. Indeed, when $\nu \to \infty$, the value of the integral associated with each of the Lefschetz thimbles converges to the geometric optics value of the field for each image given by Eq.~\eqref{eq:eik_imag}. Since Picard-Lefschetz theory gives an exact value of the integral for all values of $\nu$, this essentially allows us to extend the image analysis of geometric optics to low frequencies.

\begin{figure}
    \centering
    \includegraphics[width=\columnwidth]{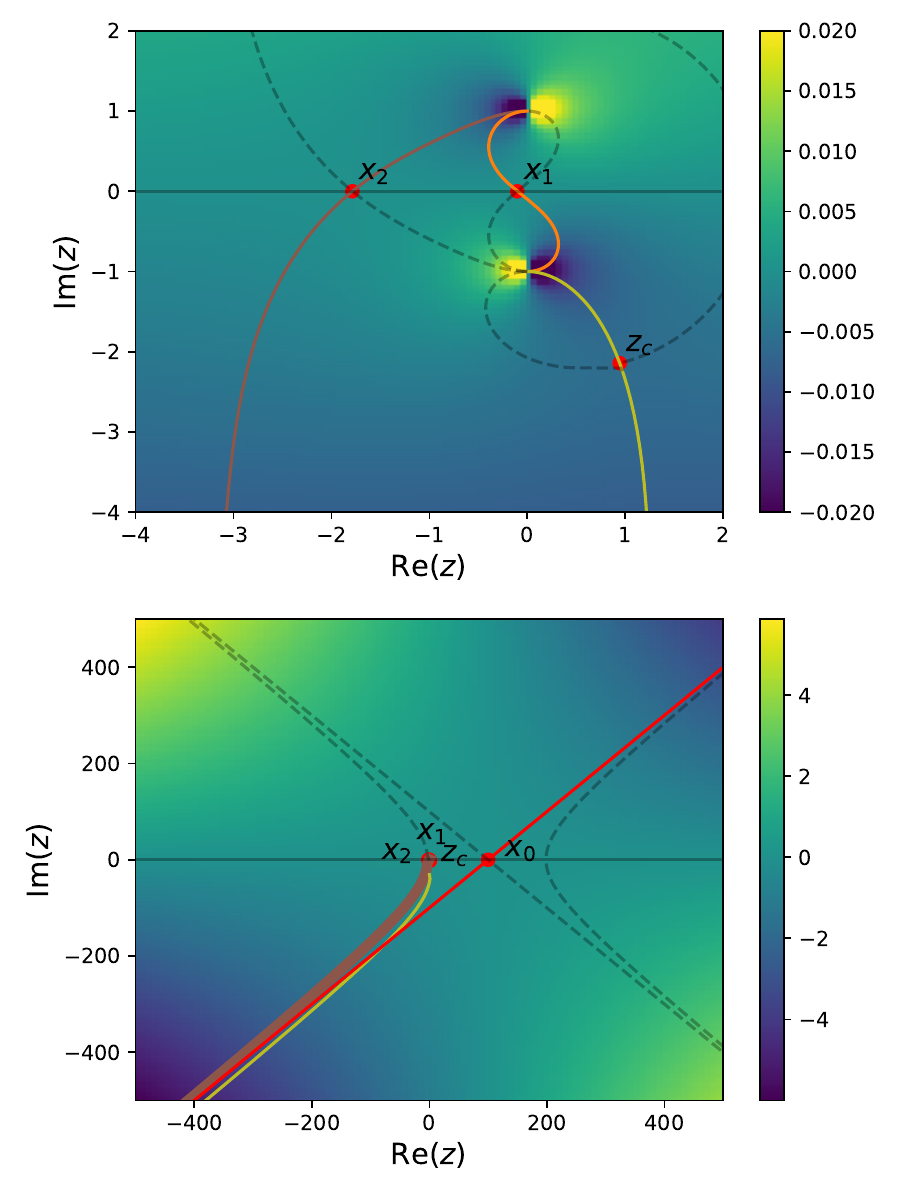}
    \caption{The Lefschetz thimbles in the complex plane for the rational lens, $\psi = 1/(1+x^2)$, with $\kappa = 500$ and $y = 100$. For these parameters there are three relevant real images, which have been labeled $x_0$, $x_1$, $x_2$, and a single relevant complex image, $z_c$. The bottom panel shows a zoomed out view of the complex plane. The background colour is determined by the value of the function $h(z)$, evaluated for $\nu = 10^{-5}$. The Lefschetz thimbles (the coloured lines) are the surfaces of steepest descent in the function $h$ from the critical points. The steepest ascent contours are shown as grey dashed lines, and the original integration contour, the real line, is shown as a solid grey line.}
    \label{fig:II_thimbles}
\end{figure}

As an example, consider when $\kappa = 500$ and $y = 100$, which corresponds to Region II in Fig.~\ref{fig:topology}. The total intensity fluctuations above unity for these parameters as a function of frequency is shown in the top middle panel of Fig.~\ref{fig:transition} (along with similar plots for the other regions). For these parameters, there are four relevant critical points, one complex and three real. We will label the complex critical point $z_c$ and the real critical points $x_0$, $x_1$, and $x_2$, with descending value. Fig.~\ref{fig:II_thimbles} shows the critical points and their corresponding Lefschetz thimbles in the complex domain. The background colour corresponds to the value of the function, $h(z)$. In general, Lefschetz thimbles terminate at poles of $h(z)$ or infinity. The thimble for $x_2$ goes from the pole at infinity to the pole at $+i$; the thimble for $x_1$ goes from the pole at $+i$ to $-i$; the thimble for $z_c$ goes from $-i$ to infinity; and the thimble for $x_0$ goes from infinity in the lower half-plane to infinity in the upper-half plane. 

\begin{figure}
    \centering
    \includegraphics[width=\columnwidth]{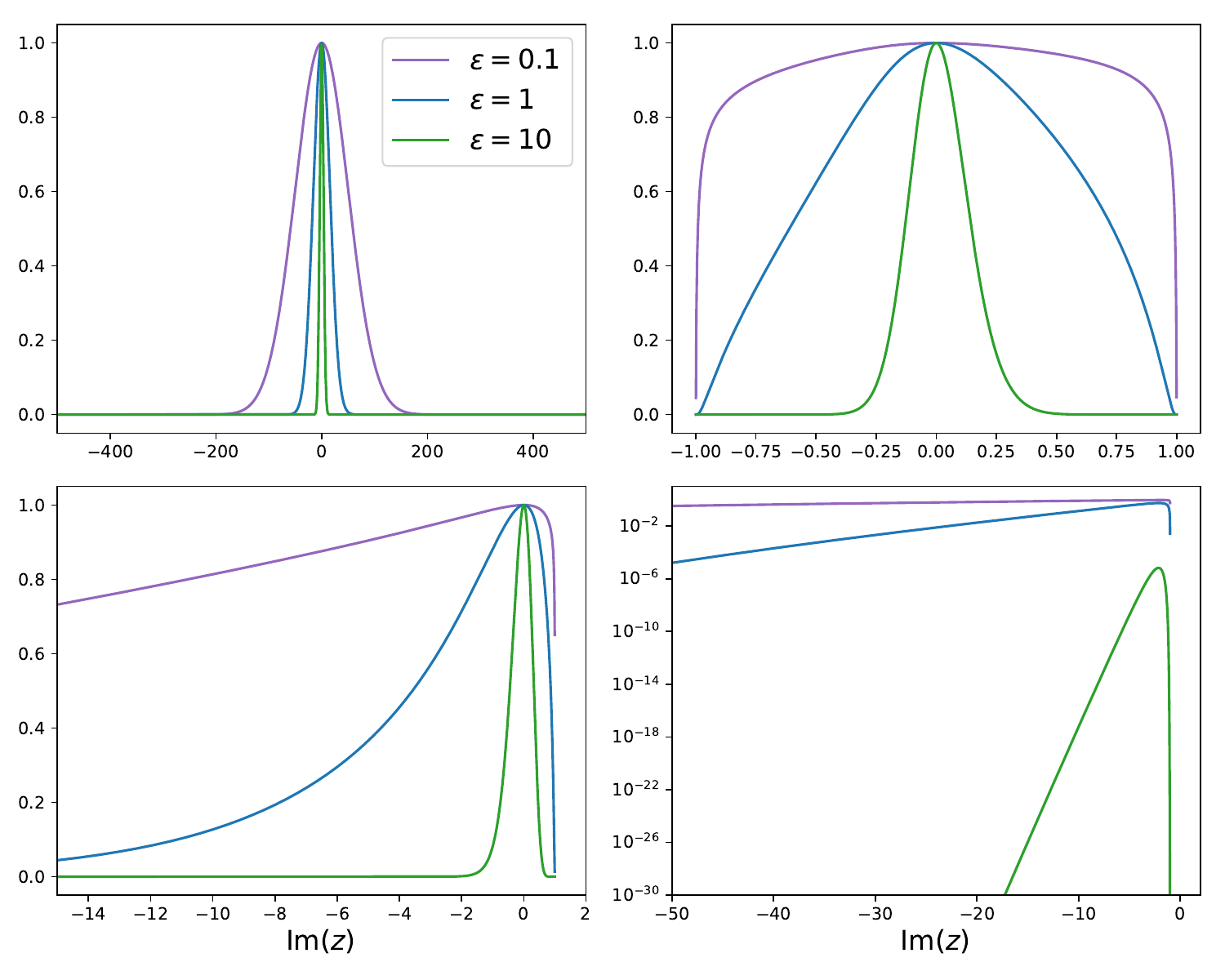}
    \caption{The absolute value of the integrand of the Kirchhoff-Fresnel integral, $\exp \{ i \nu [ \frac{(x-y)^2}{2} + \frac{\kappa}{1+x^2} ] \}$, evaluated along the Lefschetz thimbles shown in Fig.~\ref{fig:II_thimbles} for $\kappa = 500$ and $y = 100$. From left to right, and top to bottom, the integrand is shown for the thimbles associated with $x_0$, $x_1$, $x_2$, and $z_c$. The thimbles are parametrized by the imaginary component of the complex variable, $\mathrm{Im}(z)$. The integrand is shown for multiple values of $\nu$, such that $\epsilon \equiv \kappa \nu = 0.1, 1,$ and 1.}
    \label{fig:II_integrands}
\end{figure}

By examining the integrand along each of these thimbles, we can see exactly how geometric optics fails at low frequencies. Fig.~\ref{fig:II_integrands} shows the magnitude of the integrand evaluated along the thimbles for different values of $\nu$. For large $\nu$, the integrand generically takes the form of a single peak at the critical point, which decays exponentially away from the peak. As $\nu$ increases, the width of this peak decreases, and the integrand effectively becomes a delta function at the critical point. Thus, in the limit of $\nu \to \infty$, the contribution of each thimble is entirely determined by a small neighbourhood of the critical point, and geometric optics is recovered. However, as $\nu$ decreases, the effective region that contributes to the integral increases. This has two main effects. Firstly, since some of the thimbles terminate at poles, rather than infinity, at low frequencies the width of the integrand has become large enough that it effectively becomes cut off by the presence of the pole before it can decay to zero (see the intergands for $z_c$, $x_1$, and $x_2$ for low frequencies in Fig.~\ref{fig:II_integrands}). Secondly, as the effective region that contributes to the integral increases, the curvature of the thimble becomes important. In the geometric limit, when the integrand is isolated to a small region about the critical point, only the local curvature of the thimble matters. In particular, the measure of the integral, $dz$, becomes $dz \propto \exp \big\{i \big[\frac{\pi}{4} -  \frac{\mathrm{arg}(\Delta_j)}{2}\big]\big\}$ at the critical point. However, as the $\nu$ decreases, the curvature of the integration contour away from the critical point becomes relevant. While the location of the poles is independent of the lens parameters, the effective width of the integrand along the thimbles is determined by both $\nu$ and $\kappa$. In particular, it is when $\epsilon = \kappa \nu \gg 1$ that the integrand becomes localized to a small region on the thimble about the critical point with constant curvature. Thus, it is in this way that the parameter $\epsilon$ sets when geometric optics holds. Note that since the thimble for $x_0$ goes from minus infinity to positive infinity in a straight line (i.e. with constant curvature), neither of these effects is relevant, and, indeed, the contribution to the total integral computed along its Lefschetz thimble is always equal to the geometric optics prediction for all $\nu$. 

As an aside, the fact that near the critical point, the measure along the Lefschetz thimble becomes  $dz \propto \exp \big\{i \big[\frac{\pi}{4} -  \frac{\mathrm{arg}(\Delta_j)}{2}\big]\big\}$ is the origin of the extra phase factors in Eqs.~\ref{eq:eik_real} and \ref{eq:eik_imag}. When the critical points are real, $\mathrm{arg}(\Delta_j) = 0$ or $\pi$. This binary choice of angle for real images is usually expressed as the statement that each real image has a Morse index, $n_j$, associated with it, which is an integer of either $0$ or $1$. However, when the formalism is extended into the complex plane, we see that for complex images, the angle associated with the Lefschetz thimble is arbitrary, and so images, in general, do not have an associated integer Morse index. 

\begin{figure}
    \centering
    \includegraphics[width=\columnwidth]{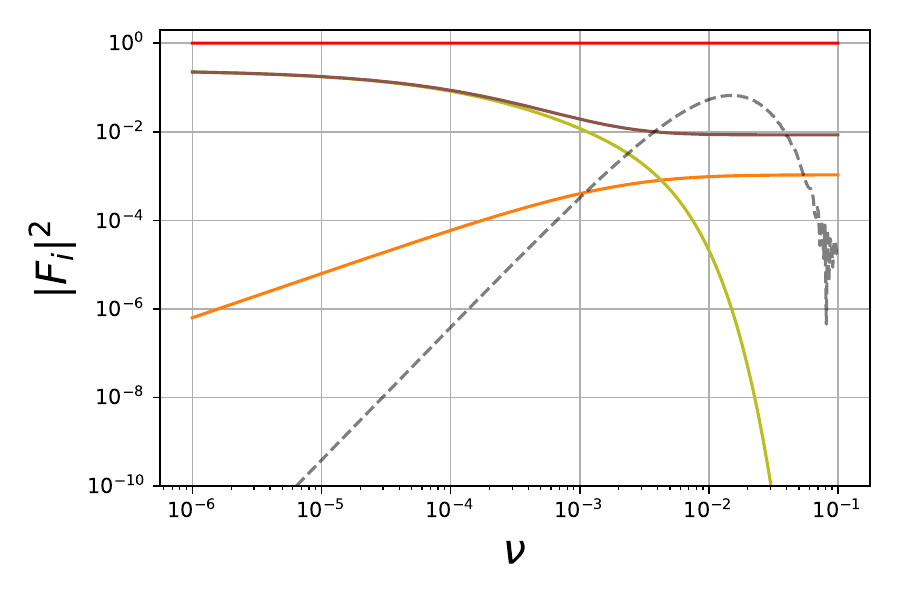}
    \caption{The magnitude squared of the contribution to the total Kirchhoff-Fresnel integral associated with the Lefschetz thimbles of $x_0$ (red), $x_1$ (orange), $x_2$ (blue), and $z_c$ (green), for the rational lens with $\kappa = 500$ and $y = 100$. The thimbles and images are shown in Fig.~\ref{fig:II_thimbles}. The grey dashed line shows the perturbative intensity fluctuations, $|F^\mathrm{pert.} - 1|^2$.}
    \label{fig:II_images}
\end{figure}

Now, examining how the integrand behaves along the Lefschetz thimbles allows us to understand how geometric optics systematically fails as $\nu$ decreases. We can also look at how the integral evaluated along the thimbles separately behaves as a function of frequency. Fig.~\ref{fig:II_images} shows the flux associated with each image as a function of frequency. That is, we compute the integral along each thimble separately and plot the absolute value squared of the result. In the high-$\nu$ limit, these values converge to the intensities $|F_j|^2$ for each of the images predicted by geometric optics. The grey dashed line in Fig.~\ref{fig:II_images} shows the value of the intensity fluctuations in the perturbative expansion, $|F^\mathrm{pert.}  - 1|^2$. We know from Fig.~\ref{fig:transition} that the total intensity computed using Picard-Lefschetz theory begins to agree with the perturbative below $\nu \sim 10^{-2}$; however, in Fig.~\ref{fig:II_images} we find that the intensities of the individual images are, asymptotically, orders of magnitude larger than the perturbative intensity fluctuations. While this may seem strange at first glance, this is not contradictory. As $\nu \to 0$, we can see that the intensities of the images at $z_c$ and $x_2$ become equal to each other. Their phases, however, become almost exactly opposite. This is easy to see from the bottom panel of Fig.~\ref{fig:II_thimbles}. At large $|z|$, the thimbles associated with $z_c$ and $x_2$ become anti-parallel. When $\nu \to 0$, the relevant region of integration along these thimbles gets larger, and so the values of the integral for these two images become roughly equal in magnitude but opposite in sign. Thus, while individually their intensities may be large, the magnitude of their sum is small. It turns out that this sum is close to opposite in phase with the image at $x_1$, leading to a further cancellation. As a result, the intensity of the sum of the three images $x_1$, $x_2$, and $z_c$ becomes very small, and, indeed, agrees with the perturbative intensity fluctuations (the intensity for the image $x_0$ is unity, independent of $\nu$). 

This gives us insight into how geometric optics becomes diffractive optics as $\nu$ decreases. In this case, a subset of the images become mutually coherent with each other, in such a way that they almost exactly cancel each other out. What is left is an unperturbed image ($x_0$ in this case) with a small modulation on top. We can perform the same exercise for the different regions shown in Fig.~\ref{fig:topology}, and a similar picture emerges: the multiple images in the geometric picture attain the same or opposite phases in the limit as $\nu \to 0$, and their sum yields the perturbative expansion. 

\bsp	
\label{lastpage}
\end{document}